\PassOptionsToPackage{table}{xcolor}
\documentclass[12pt]{article}
\usepackage[T1]{fontenc}
\usepackage[utf8]{inputenc}
\usepackage[mathcal]{euscript}
\usepackage{amsmath, amssymb, amsthm}
\usepackage[doublespacing]{setspace}
\usepackage[left=1.00in, right=1.00in, top=1.00in, bottom=1.00in]{geometry}
\usepackage{enumitem}
\usepackage{xcolor}
\usepackage{soul}
\usepackage{pdfpages}
\usepackage{bbm}
 \usepackage{fancyhdr}
 \usepackage{graphicx}
\usepackage{subcaption}
\usepackage{tikz}
\usepackage{pgfplots}
\usepackage{floatrow}
\usepackage{hyperref}
\usepackage{appendix}
\usepackage[round]{natbib}
\usepackage{booktabs}
\usepackage{siunitx}
\usepackage{comment}
\usepackage{pdflscape}
\newcolumntype{d}{S[input-symbols = ()]}
\usepackage[para,online,flushleft]{threeparttable}
\pgfplotsset{compat=1.17} 
\usepackage[export]{adjustbox}
\usepackage{titling}    
\title{High-Dimensional Spatial-Plus-Vertical Price Relationships and Price Transmission: \\ A Machine Learning Approach} 
\date{\today}

\begin{document}

\begingroup
  \singlespacing       
  \maketitle
\endgroup

\begin{singlespace}
\begin{center}
\textbf{Mindy L.\ Mallory} \footnote{The authors would like to thank Bernhard Dalheimer for helpful comments to this paper. }\\
Department of Agricultural Economics, Purdue University\\
\textit{Corresponding author:} \href{mailto:mlmallor@purdue.edu}{mlmallor@purdue.edu}
\end{center}

\begin{center}
\textbf{Rundong Peng}\\
Department of Agricultural, Food, and Resource Economics,\\
Michigan State University
\end{center}

\begin{center}
\textbf{Meilin Ma}\\
Department of Agricultural Economics, Purdue University
\end{center}

\begin{center}
\textbf{H.\ Holly Wang}\\
Department of Agricultural, Food, and Resource Economics,\\
Michigan State University
\end{center}

\vspace{1.2em}

\end{singlespace}

\clearpage        

\begingroup
  \singlespacing        
  \maketitle
\endgroup

\doublespacing    

\thispagestyle{empty}
\begin{abstract} 
\singlespacing
Price transmission has been studied extensively in agricultural economics through the lens of spatial and vertical price relationships. Classical time-series econometric techniques suffer from the “curse of dimensionality” and are applied almost exclusively to small sets of price series --- either prices of one commodity in a few regions or prices of a few commodities in one region. However, an agrifood supply chain usually contains several commodities (e.g., cattle and beef) and spans numerous regions. Failing to jointly examine multi-region, multi-commodity price relationships limits researchers' ability to derive insights from increasingly high-dimensional price datasets of agrifood supply chains. We apply a machine-learning method -- specifically, regularized regression -- to augment the classical vector error correction model (VECM) and study large \textit{spatial-plus-vertical} price systems. Leveraging weekly provincial-level data on the piglet-hog-pork supply chain in China, we uncover economically interesting changes in price relationships in the system before and after the outbreak of a major hog disease. To quantify price transmission in the large system, we rely on the \textit{spatial-plus-vertical} price relationships identified by the regularized VECM to visualize comprehensive spatial and vertical price transmission of hypothetical shocks through joint impulse response functions. Price transmission shows considerable heterogeneity across regions and commodities as the VECM outcomes imply and display different dynamics over time. 
\end{abstract}

\textit{JEL Codes:} Q02, Q11, Q18

\textit{Keywords:} spatial and vertical price relationships, price transmission, high-dimensional data, machine learning

\pagebreak 
\clearpage
\pagenumbering{arabic} 
\doublespacing

\section{Introduction}
Price transmission refers to the extent to which shocks in one market generate price effects elsewhere and has been studied extensively in agricultural economics through the lens of spatial and vertical links: relationships among prices of the same commodity across space and along the supply chain for vertically related commodities \citep{chavas2020dynamics, von2021price}. In the agricultural sector, data have become available for many supply chains that contain several vertical stages, and each stage operates in numerous regional markets. The high-dimensional system implies complex \textit{spatial-plus-vertical} price relationships that drive price transmission. As a result, when the price of one commodity (e.g., hog) changes, it would not only transmit to other regional markets of the same commodity, but also to markets of related commodities (e.g., pork) in home as well as other regions. 

However, classical time-series econometric techniques, including vector auto-regression (VAR) and vector error correction model (VECM), suffer from the 'curse of dimensionality' --- as the number of parameters increases at a rate of at least $m^2$ with the number of variables, $m$. As a result, VAR and VECM have been applied almost exclusively to small sets of price series, either the prices of a commodity in a few regions or the prices of a few commodities within a region \citep{fackler2001, wang2023dynamic}, missing potential insights into spatial-plus-vertical price relationships in the larger system. 

We follow recent advances in machine learning (ML) methods on estimating sparse cointegrating systems using regularized regression and fit a high-dimensional VECM to a set of vertically related commodities and across a large number of regions. Using data on the Chinese piglet-hog-pork industry, we estimate the VECM of spatial-plus-vertical network of prices based on regularized regressions and reveal complex price relationships that change after a major disruption hit the supply chain, the African Swine Fever (ASF) epidemic in 2018 that wiped out a quarter of sow and hog stocks in China \citep{ma2021african}. During the shock, inter-province shipments of live hogs were banned for several months, breaking down the spatial integration in the industry \citep{ma2023risk}. 

Rich spatial and vertical price relationships are uncovered by the high-dimensional VECM. We find evidence of structural breaks punctuated by the imposition and removal of the shipping ban in the price relationships.  The changes appear most pronounced in the relationships governing the hog market, the market most directly affected by the ASF outbreak. The reduced-form long-run relationships impacting hog prices become more concentrated, and short-run dynamics are weakened.\footnote{The findings of more concentrated long-run relationships are consistent with investments made by the Chinese government in increasing the concentration of hog production and processing. See more discussion in Section \ref{sect:result}.}

The regularized, high-dimensional VECM results produce a large number of coefficients, making interpretation a challenge for experts and general audiences alike. We hence employ the VECM estimates to inform joint impulse response functions (JIRFs) as a method to quantify price transmissions in large systems similar to impulse response functions (IRFs) from traditional structural VARs (SVARs) \citep{wiesen2024joint}. The JIRF utilizes reduced-form errors from the VAR to generate a unique propagation of shocks on a subset of the variables in the large system, making it ideal to investigate scenarios where shocks may impact many variables at once (as was the case with the ASF shock to the pork complex in China). Using reduced-form errors also sidesteps the complication of identification that limits traditional SVARs. 

Although causal interpretation is sacrificed, the JIRF allows for the exploration of price transmission through the contemporaneous correlations and cross-equation relationships revealed by the VECM. While the estimated marginal effects may be biased in the VECM estimated with regularized regression, the coefficient estimates are consistent \citep{basu2015regularized}. Crucially, the improvements in forecast variance gained from regularization methods allow us to measure price transmission from shocks to the large spatial-plus-vertical system of prices in a way that is not possible with traditional econometric approaches.  

By imposing hypothetical shocks on the piglet, hog, and pork prices, respectively, JIRF outcomes confirm significant spatial and vertical heterogeneity in price transmissions as implied by the VECM outcomes. We find that hog markets, which conduct the most inter-province trade, have the most spatially similar responses to shocks, and the responses decay at similar rates. Pork and piglet markets, which do not engage in as much inter-province trade, display larger degrees of spatial heterogeneity to responses of any given shock.  Additionally, we find that shocks transmit up and down the supply chain. Shocks to the hog market, for example, generate significant and long-lasting responses in piglet markets. 

We make three contributions to the literature. First, we show how regularized regression can be used to fit models of high-dimensional systems that are not possible with traditional regression techniques. This is important for researchers interested in uncovering complex relationships given increasingly rich data on agrifood supply chains. For example, our approach is agnostic to the spatial structure, so different spatial relationships than those implied by a spatial autoregressive model can be revealed by the data \citep{anselin2022spatial}. 

Second, we show how JIRFs can complement ML methods in a time series setting to enhance the interpretability of the estimated ML model. Some ML models of high-dimensional systems can produce a dizzying number of coefficients and marginal effects, making it hard to see the forest for the trees. Other ML models solely focusing on minimizing prediction errors are complete black boxes, leaving the researchers unable to get insights into how the model produces predictions and unable to determine statistical significance of the estimates. JIRFs allow for economically meaningful shocks and visualize how the model predicts variables of interest would respond with corresponding statistical significance. Thus, JIRFs in combination with ML methods of estimation form a powerful method for economic analysis of high-dimensional price transmission for a wide array of contexts. 

Third, this interpretability produced by the JIRFs from a high-dimensional model is a great benefit to the general audience --- practitioners, policymakers, and stakeholders. Being able to shock variables one-by-one or in economically meaningful subsets provides valuable flexibility in preparing for potential shocks in supply chains, exploring policy 'what if' analysis, or identifying nodes that make the entire system vulnerable to catastrophic shocks. Such ability is critical for agrifood and other industries because of increasingly frequent disruptions in supply chains in recent years \citep{baldwin2022risks, hadachek2024market}.

\subsection{Related Literature}
Price transmission has been studied extensively in economics and agricultural economics via two dimensions: spatial price transmission and vertical price transmission \citep{von2021price}. 

There is a long strand of literature on the Law of One Price (LOP) across regional markets using time-series and spatial econometric models \citep{fackler2001}. The LOP suggests that profit-maximizing traders would facilitate long-run equilibrium prices that differ no more than spatial arbitrage costs, including transport costs, across sub-markets. If the LOP holds, the spatially dispersed markets form an integrated economic market. 

Spatial integration has been studied, for instance, in the context of the US cattle market because declining volumes traded through cash markets generated concerns that packers with market power could distort prices in specific regions. As a result, spatial integration may weaken \citep{goodwin1991cointegration, pendell2006impact}.  Economic reform and market liberalization in many developing economies since the 1990s have also inspired a large number of studies on the dynamics of spatial integration with a focus on evaluating the effect of development policies \citep{badiane1998, abdulai2000}. Employing tests for spatial integration based mainly on time-series price data such as Baulch’s parity bounds integration models, regime switching models, and threshold VECM, most have found evidence of positive policy impacts on market integration. 

How market prices evolve vertically along a supply chain has key implications on producer and consumer welfare. Low pass-through rates of cost decreases from producers to consumers, for instance, can translate into larger markup for retailers, which harms consumers, and may reflect the exercise of seller power. 

A key issue to investigate is the nature of vertical price transmission, including its magnitude, speed, and symmetry, under demand and supply shocks \citep{peltzman2000prices, chavas2020dynamics}. Many empirical studies examine the vertical price transmission in contexts of various agricultural commodities, obtaining mixed findings. For example, \citet{brummer2009impact} estimate the long-run elasticity of price transmission between wheat and flour prices in the Ukraine to be 0.81. Using an asymmetric error correction model, \citet{von1998estimating} studies the transmission between farmer and wholesaler pork prices in Germany and finds asymmetric price transmission. \citet{abdulai2002using} employs threshold cointegration tests and documents asymmetric price transmission between producer and retailer in the Swiss pork market. \citet{miller2001price} and \citet{assefa2017price} also find evidence for asymmetric price transmission along the pork supply chain. In contrast, \citet{bakucs2005marketing} find symmetric transmission between farm hog price and retail pork price in the Hungarian market. None of these vertical market studies, though, consider the spatial markets at the same time. 

Conventional spatial analysis only reveals how prices of a commodity across local markets relate to each other, and a conventional vertical analysis reveals how prices of several commodities relate to one another within a local market. For one commodity in a supply chain, the price impact of related commodities in other markets is, therefore, not captured. For example, \citet{wang2023dynamic} employ 7 region-level (i.e., each region consists of multiple provinces) price series for hog and pork, respectively, employing 14 series in total. Their spatial price analysis is conducted across pairs of regions, while their vertical price analysis is conducted by region. Bivariate relationships have been used to examine large complex systems of prices in other markets as well, see \citet{goodwin1991cointegration} and \citet{pendell2006impact} for examples.

\section{Data}
The 2018 ASF outbreak, a highly contagious and deadly disease for hogs, resulted in substantial losses of hogs and sows in China, the world’s largest hog/pork producer and consumer. China has a self-sufficiency rate for pork that is greater than 96\% and produces 5.0-6.0 million tons of pork per year during the period of interest \citep{ma2021african}. China’s pork consumption is concentrated in large coastal cities, not coinciding geographically with inland production regions due to land, labor, and environmental constraints. Without advanced cold chain transportation, inter-province transportation of live hogs played a key role prior to ASF.

Soon after the ASF outbreak in August 2018, the Chinese government banned inter-province live hog shipments to prevent the ASF virus from spreading across provinces; a typical policy reaction facing animal epidemics. Not surprisingly, the once closely co-moving provincial hog prices diverged significantly under the bans. By March 2019, most of the bans were lifted, hog prices started co-moving more closely \citep{ma2023risk}.\footnote{The ASF outbreak was a surprise to all. There is no evidence that the ASF outbreak time depends on the intensity of provincial hog trade, production, or consumption. Almost immediately after the first ASF case was confirmed, bans rolled out across provinces, leaving the start time for the ASF period clear and exogenous. The end time of the ASF period is also exogenous and clear because most provinces successfully cleared ASF cases within their boundaries and had their bans removed by March 2019.}


We collect weekly price data for piglets, hogs, and pork for Chinese provinces, covering September 27, 2016 to January 10, 2023 (298 weeks in total). The original data are daily and county specific. We convert the data to the province-week level by simple averaging. Out of all mainland provinces, Hainan, Ningxia, Qinghai, and Tibet are excluded from our sample because of relatively large numbers of missing observations during the period of interest. Given that they are all small producers of hogs and small consumers of pork (see Figure \ref{fig:hogout}), we are not concerned about excluding them from the empirical analysis. There are small numbers of missing observations for the remaining 27 provinces in the sample. Because the missing observations are usually isolated and infrequent, we use simple linear interpolation to complete the dataset.\footnote{Among the price series, there are 533 missing observations (week-province pairs) in piglet, 56 missing observations in hog, and 504 missing observations in pork. The number of missing observations is below 7.0 percent for both piglet and pork price series and below 1.0 percent for hogs. See Table \ref{tab:stat} for more information.}

\begin{table}[H]
\centering\caption{Summary Statistics}
\begin{threeparttable} 
\footnotesize


\begin{tabular}{lccccccc}
\hline
 & Mean & Std. Dev. & Min  & Median & Max & Unit & Missing data \%\\ 
\hline
\textit{Pre Period} \\
Province piglet price & 35.00 & 9.43 & 13.84 & 33.13 & 74.62 & RMB/kg & 5.99  \\ 
Province hog price & 14.32 & 2.37 & 9.06 & 14.34 & 23.60 & RMB/kg & 0.59  \\ 
Province pork price & 20.89 & 3.53 & 10.61 & 20.78 & 31.27 & RMB/kg & 6.91  \\
\textit{Post Period 1} \\
Province piglet price & 81.47 & 29.98 & 18.85 & 81.89 & 261.58 & RMB/kg & 2.54  \\ 
Province hog price  & 26.90 & 7.39 & 10.39 & 28.91 & 41.25 & RMB/kg & 0.64  \\ 
Province pork price  & 36.47 & 9.73 & 13.54 & 38.73 & 62.33 & RMB/kg & 1.60  \\
\textit{Post Period 2} \\
Province piglet price  & 32.21 & 10.18 & 10.76 & 31.58 & 73.33 & RMB/kg & 1.96  \\ 
Province hog price  & 15.98 & 4.07 & 9.01 & 14.80 & 26.20 & RMB/kg & 0.00  \\ 
Province pork price  & 22.78 & 5.09 & 11.29 & 21.66 & 40.38 & RMB/kg & 0.22  \\
\hline
\end{tabular}

\begin{tablenotes}
\footnotesize{\textit{Source}: \url{http://www.zhujiage.com.cn/}.

\textit{Note}: Price data are deflated using a common factor, China’s monthly Consumer Price Index (\url{http://www.stats.gov.cn}), with January 2018 as the baseline. The prices are initially reported at the county-day level. Data are aggregated to the province level and weekly basis by simple averaging. \textit{Pre} Period refers to the time before the ASF shipping ban was in place (September 26, 2016 to August 30, 2018), \textit{Post1} refers to the short-term after the shipping ban was lifted (March 20, 2019 to July 4, 2021), and \textit{Post2} refers to the rest of the series (July 5, 2021 to January 10, 2023).}
\end{tablenotes}
\end{threeparttable}
\label{tab:stat}
\end{table}

Table \ref{tab:stat} reports the summary statistics of the finalized dataset. All prices are measured in real RMB per kilogram, using China's monthly Consumer Price Index with a January-2018 baseline. Piglet has the highest mean prices and shows higher price volatility than hog and pork. The mean hog price is the lowest compared to the piglet and pork prices. Pork prices show higher volatility than hog prices.

\begin{figure}[H]
    \centering 
    \includegraphics[width=0.9\textwidth]{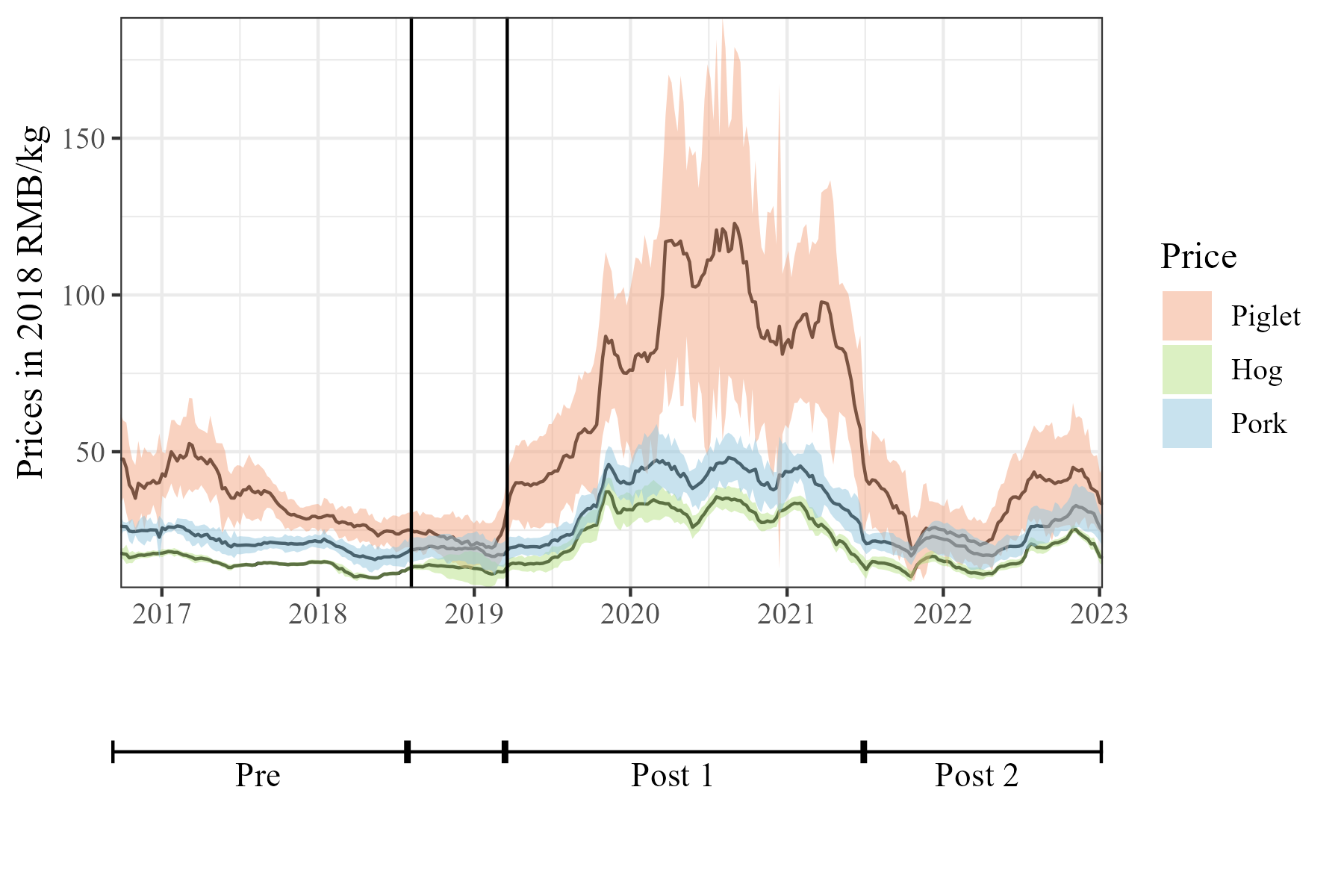}
    \caption{Weekly Nation-Level Prices of Piglet, Hog, and Pork (2016-2023)
    \smallskip \\ \footnotesize \textit{Note}: The color band represents the two-standard-deviation interval around the national average price, captured by the solid line, for 27 Chinese provinces in the sample. Each point on the upper (lower) bound of the band represents the average price plus (minus) two standard deviations. The horizontal axis represents weeks from September 26, 2017 to January 10, 2023. From left to right, the two vertical lines indicate the start of the ASF shipping ban on August 31, 2018 and the removal of the ban on March 19, 2019, respectively. The period after the ban is split into two periods (\textit{Post1} and \textit{Post2}) at July 5, 2021.} 
    \label{fig:meansd}
\end{figure}

Figure \ref{fig:meansd} shows the national average prices of piglet, hog, and pork prices with two-standard-deviation bands (average and standard deviation calculations across provinces for a given point in time). The period of the shipping ban on hogs and pork is shown by the two vertical bars. The \textit{Pre}, \textit{Post1}, and \textit{Post2} periods identified in the lower portion of the figure are the result of the Chow test for structural breaks in our model (see Section \ref{sect:result} for details). 

A few patterns stand out by visualizing the data. First, the shipping ban imposed to slow the spread of ASF results in considerable divergence in price movements in the three markets, which is visible in a widening of the bands during and after the shipping ban. Second, the upstream piglet market and downstream pork market are heavily affected, although pork or piglet shipping was not banned. Third, the three stages demonstrate differential patterns in the divergence and reintegration of provincial prices.

\section{Empirical Model}\label{sect:empiric}   
We employ price series for 27 provinces and for piglet, hog, and pork, totaling 81 series. The goal is to model the spatial and vertical price relationships in the entire network of prices to examine all the spatial-plus-vertical links, using ML regularization methods to overcome the overfitting and curse of dimensionality challenges of estimating such a large and complex system. 

VECMs estimated using maximum likelihood are notoriously problematic to fit for large systems due to high requirement on number of observations. \citet{johansen1995likelihood} provides critical values for testing cointegrating rank up to eleven series. However, Johansen's likelihood test performs poorly for more than four or five series, or when the model is mis-specified \citep{mallory2012testing}. Therefore, estimating our system of 81 price series with standard methods is not suitable. 

The penalized least-squares method is able to estimate a large number of parameters of the VECM, even for a modest sample size of observations over time. Table \ref{tab:workflow} previews the steps we follow in conducting the analysis. The steps are described in more detail in the following subsections, and a set of variables are introduced. Table \ref{tab:glossary} serves as a notation reference for the reader.

\begin{table}[!ht]
    \centering\caption{Workflow of Empirical Analysis}
\begin{threeparttable} 
\footnotesize
\begin{tabular}{p{0.2\linewidth} p{0.7\linewidth}}
\hline
\textbf{Workflow} & \textbf{Description} \\
\hline
Exploratory data analysis & Calculate summary statistics for piglet, hog, and pork prices. \\
\\
Transform data & Convert prices to logarithmic scale.\\
\\
Unit root & Check for stationarity in the log-transformed price data with panel unit root test. \\
\\
VAR model estimation & Estimate VAR model in levels with penalized least squares. \\
\\
VECM model and $\Pi$ & Recover VECM from the estimated VAR in levels (Granger Representation Theorem) and obtain $\Pi$ matrix. \\
\\
Cointegration and the effective rank of $\Pi$ & Carry out Roy and Vetterli (2007) Effective Rank test of the rank of $\Pi$ to confirm it is reduced rank in \textit{Pre}, \textit{Post1}, \textit{Post2}. \\
\\

VAR to VMA & Transform the VAR to a Vector Moving Average (VMA) process to get the coefficient matrices to compute the JIRF. \\ \\
JIRF & Use bootstrapped VMA to compute the mean and 95\% confidence interval of the JIRF. 
\\
\hline
\end{tabular}

\begin{tablenotes}
\item \footnotesize{\textit{Note}: This table offers an abbreviated outline of our empirical workflow.}
\end{tablenotes}
\end{threeparttable}
\label{tab:workflow}
\end{table}

\begin{table}[!ht]
    \centering\caption{Glossary of Variables}
\begin{threeparttable} 
\footnotesize
\begin{tabular}{ll}
\hline
\textbf{Symbol} & \textbf{Description} \\
\hline
$\mathbf{Y}_t$ & $m$-dimensional multivariate time series vector at time $t$ \\
$y_{t, i}$ & The $i^{th}$ element of the multivariate time series vector $\mathbf{Y}_t$ \\
$\mathbf{c}$ & $m \times 1$ vector of intercept terms \\
$\mathbf{\Phi}_k$ & $m \times m$ matrices of coefficients for lag $k$ \\
$\boldsymbol{\epsilon}_{t}$ & Multivariate $iid$ error vector at time $t$ with mean $\mathbf{0}_m$ and covariance matrix $\Sigma$ \\
$\Delta \mathbf{Y}_t$ & First difference of the multivariate time series vector $\mathbf{Y}_t$ \\
$\boldsymbol{\Pi}$ & $m \times m$ matrix of long-run relationships in the VECM \\
$\boldsymbol{\Gamma}_i$ & $m \times m$ matrices of short-run dynamics in the VECM for lag $i$ \\
$\boldsymbol{\alpha}$ & $m \times r$ matrix of speed of adjustment parameters in the VECM \\
$\boldsymbol{\beta}$ & $r \times m$ matrix of cointegrating vectors defining long-run equilibria in the VECM \\
$\boldsymbol{\Theta}$ & $m \times (mp+1)$ parameter matrix containing intercept and coefficient matrices in block form \\
$\|\mathbf{x}\|_2$ & Euclidean (or $\ell_2$) norm of a vector $\mathbf{x}$ \\
$\|\mathbf{x}\|_1$ & $\ell_1$ norm (sum of absolute values) of a vector $\mathbf{x}$ \\
$\|\mathbf{A}\|_F$ & Frobenius norm of a matrix $\mathbf{A}$ \\
$\hat{\boldsymbol{\Theta}}$ & Penalized least squares estimates of $\boldsymbol{\Theta}$ \\
$\lambda$ & Tuning parameter in the penalized least squares estimation \\
$\gamma$ & Mixing parameter between Ridge and Lasso penalties in the penalized least squares estimation \\
$\sigma$ & Vector containing the singular values of a matrix \\
$\text{erank}(\cdot)$ & Effective rank of a matrix \\
H & Horizon of impulse responses\\
\hline
\end{tabular}

\begin{tablenotes}
\item \footnotesize{\textit{Note}: Glossary of variables introduced in the following subsections.}
\end{tablenotes}
\end{threeparttable}
\label{tab:glossary}
\end{table}

\subsection{VAR and VECM Representations of Cointegrated Time Series} 
The VAR and VECM are workhorses of applied time series analysis, so we introduce the models only briefly here. A complete and approachable treatment can be found in \citet{hunter2017multivariate}. 

Consider an $m$-dimensional multivariate time series $\left\{\mathbf{Y}_t \right\}$, so that $\mathbf{Y}_{t}=\left(y_{t, 1}, \ldots, y_{t, m}\right)'$  are  $m \times 1$ vectors on which a VAR model of order $p$ ($\operatorname{VAR}(p)$) can be defined as follows. 
\begin{equation}
    \mathbf{Y}_{t}=\mathbf{c} + \mathbf{\Phi}_{1} \mathbf{Y}_{t-1} + \mathbf{\Phi}_{2} \mathbf{Y}_{t-2} + \dots + \mathbf{\Phi}_{p} \mathbf{Y}_{t-p} + \boldsymbol{\epsilon}_{t}, 
\label{eq:varp}
\end{equation}
where $\mathbf{c} = \left(c_{1}, \ldots, c_{m}\right)'$ is a vector of intercept terms, the $\mathbf{\Phi}_k$ are $m \times m$ matrices of coefficients, and $\boldsymbol{\epsilon}_{t}=\left(\epsilon_{t, 1}, \ldots, \epsilon_{t, m}\right)'$ with the $\boldsymbol{\epsilon_{t}}$ being multivariate $iid\left(\mathbf{0_m,\Sigma} \right)$. If the series in $\mathbf{Y}_t$ are $I\left(1\right)$ with at least one common trend, the series are cointegrated and the $\operatorname{VAR}(p)$ model defined in equation \ref{eq:varp} is covariance stationary \citep{johansen1995likelihood}.

The VECM form of equation \ref{eq:varp} can be expressed as follows with $\Delta \mathbf{Y}_t$ being the first difference of $\mathbf{Y}_{t}$.
\begin{equation}
    \Delta \mathbf{Y}_t = \mathbf{c} + \boldsymbol{\Pi}\mathbf{Y}_{t-1} + \boldsymbol{\Gamma}_1 \Delta \mathbf{Y}_{t-1} + \boldsymbol{\Gamma}_2 \Delta\mathbf{Y}_{t-2} + \dots + \boldsymbol{\Gamma}_p \Delta \mathbf{Y}_{t-(p-1)} + \boldsymbol{\epsilon}_{t}, 
    \label{eq:vecm}
\end{equation}
where $\boldsymbol{\Pi} = \left(\mathbf{\Phi}_{1} + \mathbf{\Phi}_{2} + \dots \mathbf{\Phi}_{p} - \mathbf{I}_m\right)$ and $\boldsymbol{\Gamma}_i = \left( - \sum^{p}_{j = i+1} \boldsymbol{\Phi}_j \right)$. Since $\mathbf{Y}_t$ are $I\left(1\right)$ and cointegrated, that means $\boldsymbol{\Pi}$ has rank of $r$, with $0 < r < m$, and there exists a factorization $\boldsymbol{\Pi} = \boldsymbol{\alpha \beta'}$, where $\boldsymbol{\alpha}$ is a $m \times r$ matrix and $\boldsymbol{\beta'}$ is a $r \times m$ matrix \citep{engle1987co}. 

When the normalization proposed by Johansen \cite{johansen1995likelihood} is used, the $\boldsymbol{\beta}$ matrices are interpreted  as cointegrating vectors defining long-run equilibria, namely, linear combinations of the series in $\mathbf{Y_t}$ that are stationary. The $\boldsymbol{\alpha}$ are interpreted as speed of adjustment parameters determining how fast each series responds to get back to the long-run equilibria after a shock. 

\subsection{Challenges in the Estimation of the Rank of $\Pi$ and the Identification of Long-Run Relationships in Large Systems }
A fundamental part of estimating VECMs is to determine the rank of $\Pi$ and learn the number of long-term equilibrium relationships present in the system of prices. However, this is challenging in high-dimensional settings. The Johansen tests for cointegrating relationships are based on a likelihood ratio test that is asymptotically $\chi^2$\citep{johansen1988statistical}. However, in a high-dimensional VECM the sample size required to rely on the asymptotic result is impractically large for most applications because the large number of coefficients to estimate is limited by the degrees of freedom. 

Once the rank of $\Pi$ is known, \citet{johansen1988statistical} provides a normalization on the $\boldsymbol{\beta'}$ matrix where the first $r \times r$ submatrix of  $\boldsymbol{\beta'}$ is set to $\boldsymbol{I}_r$ (i.e., the $r \times r$ identity matrix). If economic theory can guide enough restrictions, other identification schemes for obtaining a unique factorization of $\Pi$ can be undertaken \citep{hunter2017multivariate, johansen1994identification, juselius1995purchasing}.  In small dimensional systems (i.e., small $m$), fitting VECMs on small systems often produce Johansen tests of cointegrating rank with $r = 1$ or $2$. The assumptions embedded in the Johansen normalization are not that heroic in this case. In the case of $r=1$, $\boldsymbol{\alpha}$ and $\boldsymbol{\beta}'$ can be identified simply by setting a single element of the $\boldsymbol{\beta}'$ to $1$, for example. Therefore, most applied economics papers with small $m$ just apply the Johansen normalization and interpret the rows of $\boldsymbol{\beta}'$ as cointegrating vectors presumably without much harm.

For large systems, in contrast, using the Johansen normalization to identify the long-run equilibrium relationships requires one to believe that there is an economic meaning in setting the first $r$ variables in the system as independent anchor points in $r$ different linearly independent equilibrium relationships. In the case of economic theory supplied restrictions, more than $r^2$ restrictions may be required because $r^2$ is the minimum required for specific patterns of restrictions. In practice, obtaining enough identifying restrictions from economic theory and proving that one has fully identified the factorization of the $\Pi$ matrix becomes increasingly difficult as the dimension of the problem grows in $m$. Fitting a VECM in high-dimensional settings, therefore, requires us to be comfortable with what information we can obtain from the estimated $\Pi$ matrix and its estimated rank, because the identification of $\boldsymbol{\alpha}$ and $\boldsymbol{\beta}'$ would usually be impossible in an economically meaningful way.

\subsection{Estimation with Penalized Least Squares}
Our application of the VECM is to a large system of prices, and ordinary least-squares estimates of the parameters in equation \ref{eq:varp} would suffer from the \textit{curse of dimensionality} since the number of parameters to be estimated grows at a rate of the squared number of series in the model. Therefore, for modest sample sizes in $t$, conventional methods would use up most of their degrees of freedom and yield parameter estimates that are imprecise. 

\citet{basu2015regularized} provide non-asymptotic upper bounds of the estimation errors of regularized estimates. They show that consistent estimation is possible using regularization methods (e.g., the elastic net regularization used in this study). 

Specifically, the elastic net regularization fits a VAR($p$) with \textit{Lasso} (least absolute shrinkage and selection operator) and Ridge penalties by minimizing an objective function that is the sum of squared residuals plus the elastic net penalty term \citep{zou2005regularization, fuleky2019macroeconomic}. For notational convenience, define $\boldsymbol{\Theta} = \begin{bmatrix}
    \mathbf{c} & \boldsymbol{\Phi}_1 & \boldsymbol{\Phi}_2 & \dots & \boldsymbol{\Phi}_p 
\end{bmatrix}$. Then $\boldsymbol{\Theta}$ is a $m \times (mp+1)$ dimensional matrix that contains the parameters of our VAR model from \ref{eq:varp} in the block matrix form. 

For any $\mathbf{x} \in \mathbb{R}^n$, let $\left\| \mathbf{x} \right\|$ be the Euclidean (or $\ell_2$) norm (i.e., the square root of the sum of squared elements of $x$), and $\|\mathbf{x}\|_1$ is the $\ell_1$ norm defined by  $\| \mathbf{x} \|_1 = \sum_{i=1}^{n} |x_i|
$. For any $m \times n$ matrix $\mathbf{A}$, $\left\| \mathbf{A} \right\|_F$ is the Frobenius norm.\footnote{$\|\mathbf{A}\|_F = \sqrt{\sum_{i=1}^{m}\sum_{j=1}^{n}|a_{ij}|^2}$, namely, the square root of the sum of the squares of all the elements of the matrix $\mathbf{A}$} 

With these definitions in place, we define our penalized least squared estimates of the $\operatorname{VAR}(p)$ model:
\begin{equation}
 \hat{\boldsymbol{\Theta}} \in \arg\min \left\{ \sum_{t=p+1}^{T} \left\| \boldsymbol{\epsilon} \right\|_F^2 + \lambda \left[ (1-\gamma) \left\| \boldsymbol{\Theta} \right\|_F^2 + \gamma \left\| \text{vec}(\boldsymbol{\Theta}) \right\|_1 \right] \right\},
\end{equation}
where $\lambda$ is a tuning parameter and $\gamma$ is a mixing parameter between the Ridge and Lasso penalties, respectively; both will be chosen via cross validation. 

Therefore, our model's parameter estimates are the values of the elements of $\boldsymbol{\Theta}$ that minimize the sum of squared errors of the regression equations, $\left\| \boldsymbol{\epsilon}_t \right\|_F^2$, plus $\lambda (1-\gamma)$ times the sum of squares of the parameter estimates, $\left\| \boldsymbol{\Theta} \right\|_F^2$ (i.e., the Ridge penalty), and plus $\lambda\alpha$ times the sum of the absolute value the parameter estimates, $\left\| \text{vec}(\boldsymbol{\Theta}) \right\|_1 $  (i.e., the Lasso penalty).\footnote{The notation $\text{vec}(\boldsymbol{\Theta})$ refers to the vectorization of the matrix $\boldsymbol{\Theta}$, which is essentially transforming a matrix to a vector by concatenating the columns of the matrix to form a one dimensional vector from the elements of the matrix. Therefore, this portion of the penalty term simply sums the absolute value of all the elements of $\Theta$, which is the Lasso penalty term.} We use package \textbf{sparcevar} for the statistical programming language \textbf{R} to estimate the model. The \textbf{R} script is available in an online supplement to this article.

\subsection{Estimating Cointegrating Rank in Large Systems}
Given that estimating a high-dimensional system likely means we cannot glean economic meaning from identifying $\boldsymbol{\alpha}$ and $\boldsymbol{\beta}'$ with the Johansen normalization anyway, we focus our efforts on determining if $0 < r < m$. Pinning down the exact rank is not important if we do not factor the $\Pi$; we only need to rule out $r=0$ and $r=m$ to ensure $\Pi$ is reduced rank the VECM is appropriate. 

If we can determine that each series is non-stationary with Augmented Dickey-Fuller tests administered to each series one-by-one and a panel unit root test confirming the non-stationarity in the panel, then we can make a strong case against $r=m$ because $r=m$ would imply that the series are jointly stationary.  Economic theory like spatial or vertical price linkages may be suggestive of ruling out $r=0$, but many applied settings are focused on determining the nature and degree of long-term price transmission or equilibrium relationships, so ruling out $r=0$ \textit{a priori} often would not be easy to defend.

Fortunately, other fields have developed methods for determining the rank of an estimated matrix. We briefly describe some of these approaches and the one we follow here. It is well known that one cannot determine the rank of an estimated coefficient matrix by calculating its rank in the usual way provided by standard linear algebra. This calculation would almost certainly be full-rank because none of the eigenvalues of the estimated matrix would be precisely zero \citep{gill1992testing}. Therefore, methods to estimate rank revolve around detecting how many eigenvalues of the estimated matrix are sufficiently different from zero. Unfortunately, similarly to the Johansen cointegration tests, most of the methods to estimate rank are likelihood ratio-type methods that require a distributional assumption on the errors of the model \citep{robin2000tests, kleibergen2006generalized}, which is not easy to justify in our context.

Our problem of estimating the rank of $\Pi$ is related to a general multivariate reduced rank regression. \citet{izenman1975reduced} shows that under certain conditions the rank of the coefficient matrix in a multivariate regression is analogous to the number of principal components. \citet{bunea2011optimal} build upon this work in a non-asymptotic setting where $m$ may be large relative to $n$ (i.e., the number of observations available). They argued that the rank of a large possibly sparse matrix can only be estimated above a certain noise threshold, and they define a measure of the \textit{effective rank} as the number of singular values of the matrix that are sufficiently large. Defining "sufficiently large" is effectively a hyperparameter that indicates the researcher's tolerance for noise. It is not clear how to best select such a hyperparameter because it is not amenable to typical methods like cross-validation of fit. 

We estimate the \textit{effective rank} of $\Pi$ using the method of \citet{roy2007effective}. This method is robust to distributional assumptions about the model and does not require the researcher to specify an acceptable tolerance level. We use this method in our analysis to determine if $r>0$. Specifically, if the Roy-Vetterli effective rank is greater than zero, we conclude that our model is indeed reduced rank and the cointegrated VAR is covariance stationary and hence the correct model specification. 

Consider the $m \times m$ dimensional estimated matrix, $\hat{\Pi}$, and let $\sigma$ be the vector containing the singular values of $\hat{\Pi}$, $\sigma = (\sigma_1, \sigma_2, \dots, \sigma_m)$). Further, let $p = (p_1, ..., p_Q)$, and $p_k = \frac{\sigma_k}{\| \mathbf{\sigma} \|_1}$. The effective rank is defined by the exponentiation of the Shannon entropy of the $p$ vector, $\text{erank}(\Pi) = exp(-\sum_{k=1}^Q p_k log(p_k))$.

\subsection{Joint Impulse Response Functions}
Because the coefficients of the VECM are not informative by themselves, researchers often employ IRFs after estimating a VECM to directly show price transmission of a shock throughout the system. Traditional IRFs require identification assumptions such as Cholesky decomposition, or identification based on economic theory. Identifying assumptions are often hard to defend, however, especially in large complex systems. 

\citet{wiesen2024joint} extend the generalized IRF developed by \cite{koop1996impulse} and \cite{pesaran1998generalized} to produce the JIRF which uses the conditional expectation of the correlated reduced-form shocks to deliver a unique impulse response, when a subset of variables are shocked in a large system of prices. The JIRF does not rely on identification assumptions. 

Additionally, since we cannot determine the statistical significance of the coefficients of the VECM when we estimate with an elastic net, we are limited our ability to directly draw conclusions about marginal effects from the model coefficients. However, since the elastic net shines in predictions, we can use JIRFs to measure price transmission.

To construct the JIRFs, we first express the VAR$(p)$ in \autoref{eq:varp} in its vector moving average (VMA) form as  
\begin{equation}
    Y_{t} = \tilde{c} + \sum_{H=0}^{\infty} A_H \epsilon_{t-H}
\label{eq:VMA}
\end{equation}
where $\tilde{c}$ is the VMA constant vector and $A_0$,  $A_1$, $A_2$, $\dots$ are the VMA coefficient matrices and $A_0 = I$ , the $m\times m$ identity matrix. Let $\boldsymbol{s}$ be the vector of impulses to the system, and we scale the impulse to one standard deviation for each variable receiving and impulse. For example, if we shock indices $j$, $k$, and $l$, then $\boldsymbol{s} = [\sqrt{\sigma_{jj}}, \sqrt{\sigma_{kk}}, \sqrt{\sigma_{ll}}]'$ , and $\boldsymbol{e}$ is a selector matrix that picks the columns $j$, $k$, and $l$. Further, let $\Sigma_{\epsilon} = E\left( \epsilon_t \epsilon_t' \right)$. 

Thus, the JIRF is 
\begin{equation}
    JIRF\left(H, \boldsymbol{s}, (Y_{t-1}, Y_{t-2}, \dots) \right) = A_H \Sigma_{\epsilon}\boldsymbol{e} \left( \boldsymbol{e}' \Sigma_{\epsilon} \boldsymbol{e}\right)^{-1} s
\label{eq:JIRF}
\end{equation}
The matrix, $\Sigma_{\epsilon}$ sometimes may be assumed to be normal. As this assumption is typically suspect in price series, we will obtain an empirical distribution of $A_{H}$, $\Sigma_{\epsilon}$, and $\boldsymbol{s}$ by bootstrapping, and thus obtain mean and confidence intervals for impulse responses. See the details of bootstrap in Appendix \ref{apx:bootstrap}.

\section{Results and Discussion}\label{sect:result}
This section presents the results of the high-dimensional VECM. We conducted a panel unit root test using Maddala-Wu's \citet{maddala1999comparative} approach, confirming the existence of unit roots in the panel with a \textit{p}-value of 0.95. We select the lag length with Akaike Information Criterion (AIC) which suggests a lag of two in the VAR in levels. Thus, we fit the VAR model with two lags for the periods defined earlier. 

We test for structural breaks at these key points. The onset of the shipping ban ends the \textit{Pre} period, while the removal of the ban begins the \textit{Post1} period. We do not consider the time during the ban as a separate period because there are not enough observations to fit the model. \textit{Post1} ends and the \textit{Post2} period begins when prices come back down to similar levels and patterns as before the ASF. The end of \textit{Pre} and beginning of \textit{Post1} are determined by the policy dates \citep{ma2023risk}. The break between \textit{Post1} and \textit{Post2} is determined by observing a change in the price behavior at that time. All breaks were statistically significant according to the Chow test. 
 
We compute the effective rank of the $\hat{\boldsymbol{\Pi}}$ matrices for \textit{Pre}, \textit{Post1}, and \textit{Post2} periods and find the effective ranks were 73.44, 71.95, and 71.78, respectively. We take this as evidence that the $\hat{\boldsymbol{\Pi}}$ matrix is reduced rank and the VECM is an acceptable specification. More detailed discussion of effective rank results by period can be found in Appendix \ref{apx:online}.

\begin{landscape}
\begin{figure}
    \centering
    \includegraphics[width=0.78\textwidth]{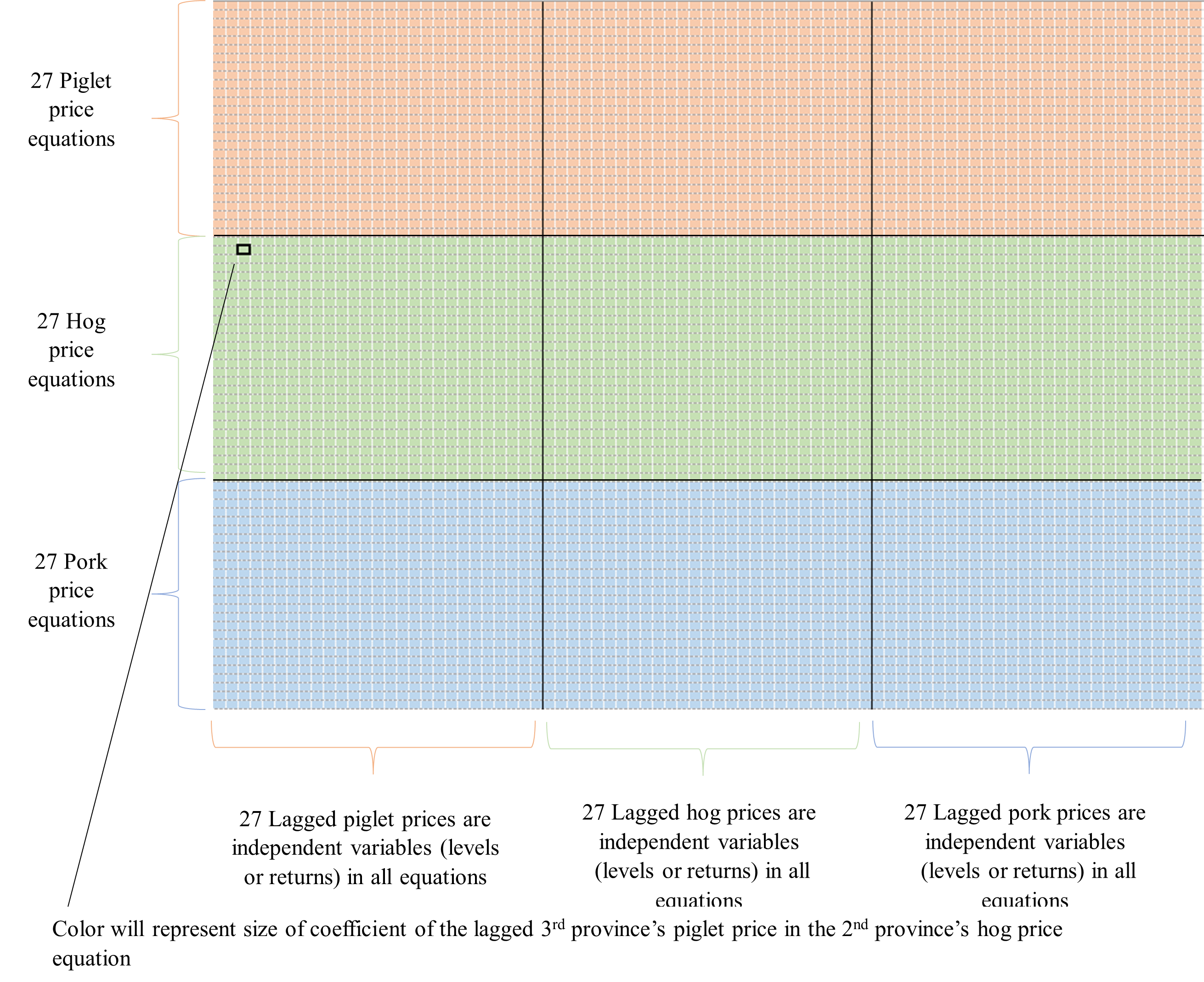}
    \caption{Example to Illustrate How VECM Results Are Presented as a Colored Grid Instead of a Table of Numbers
     \smallskip \\ \footnotesize \textit{Note}: The system is too large to examine each regression coefficient in a table. Patterns can be deduced from a colored grid where the brightness corresponds to larger magnitude estimated coefficients.}
    \label{fig:eqn-example}
\end{figure}
\end{landscape}

\subsection{Presentation of VECM Results}
Presenting results from such a large VECM system is a challenge. It is not instructive nor practical to examine and discuss every estimated coefficient. Instead, we present colored grids that correspond to the estimated coefficient $81 \times 81$ matrix where 81 equals the number of provinces (i.e., 27) times the number of commodities (i.e., 3) in our dataset. 

Figure \ref{fig:eqn-example} illustrates how the coefficients are organized. The rows are organized in three $27\times 81$ chunks where each chunk contains the piglet, hog, and pork price equations, respectively. Then, the first 27 columns with 81 rows show the lagged piglet prices' impacts in each equation, the second 27 columns show the lagged hog prices' impacts in each equation, and the third chunk of 27 columns represent the lagged pork prices' impacts in each equation. There are nine $27 \times 27$ sub-matrices bounded by the black bars. Each diagonal sub-matrix (e.g., the upper left corner sub-matrix) captures spatial price relationships for a given commodity. For a horizontal chunk (e.g., the pink section), the off-diagonal sub-matrices capture the vertical and spatial price impacts of lagged prices of other commodities in other provinces.  For example, the highlighted cell in Figure \ref{fig:eqn-example} represents the coefficient of the lagged third province's piglet price in the second province's hog price equation.

\subsection{Long-Run Relationships}
\begin{figure}[!ht]
    \centering 
    \includegraphics[width=0.9\textwidth]{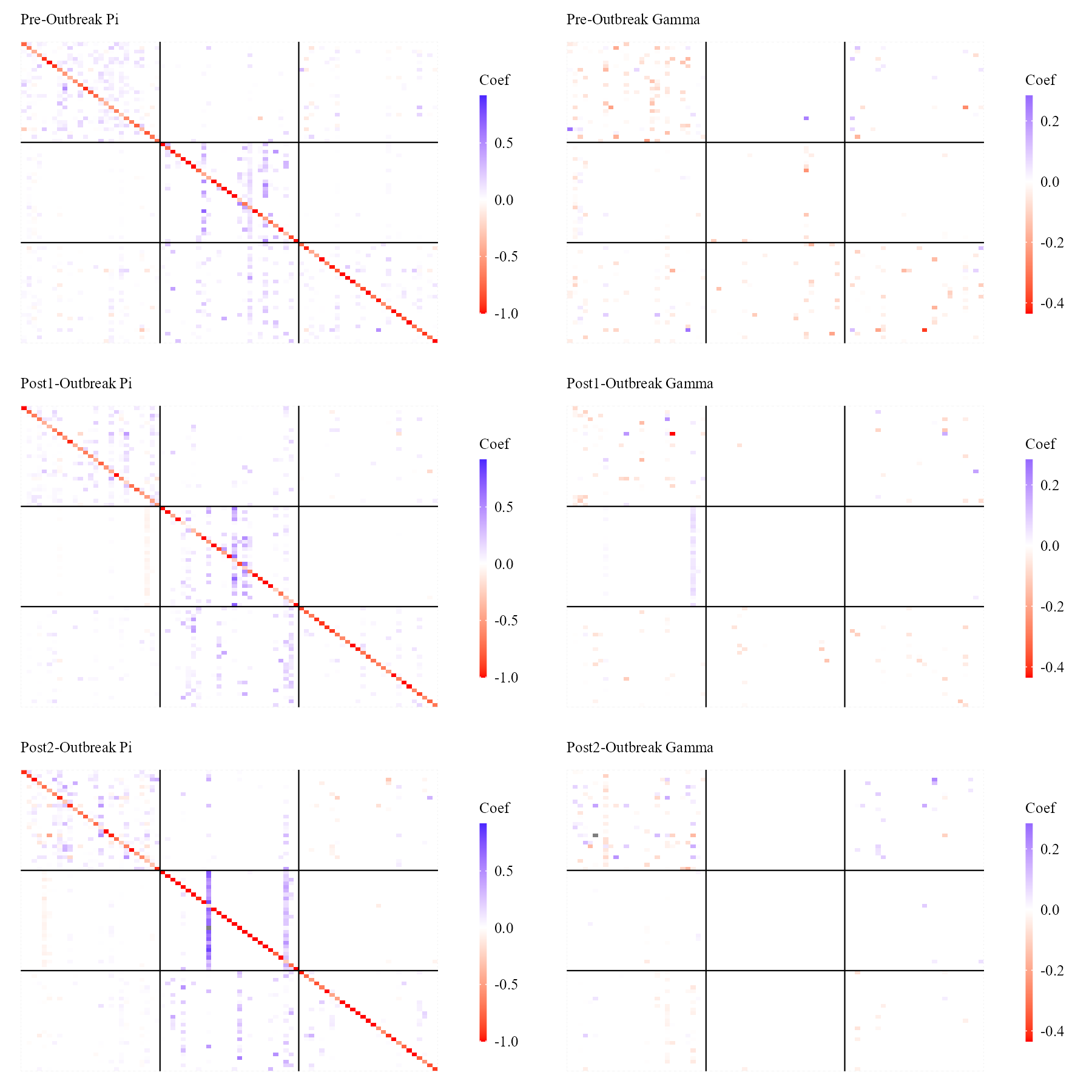}
    \caption{Parameter Estimates of VECM Model, Equation \ref{eq:vecm}
    \smallskip \\ \footnotesize \textit{Note}: Periods include the \textit{Pre} Period that refers to the time before the ASF shipping ban was in place (September 26, 2017 to August 30, 2018), \textit{Post1} that refers to the short-term after the shipping ban was lifted (March 20, 2019 to July 4, 2021), and \textit{Post2} that refers to the rest of the series (July 5, 2021 to January 10, 2023). No model was estimated during the outbreak (August 31, 2018 to March 19, 2019) since the time period is short and involves frequent policy changes.} 
    \label{fig:var-coef}
\end{figure}

Figure \ref{fig:var-coef} displays VECM outcomes for each period. The left panel presents coefficients on long-run relationships, while the right panel presents short-run coefficients. Each matrix of the coefficients are presented with province-price labels in appendix \ref{apx:online}. 

We first consider $\hat{\Pi}$'s in the left panel of Figure \ref{fig:var-coef}. Several prominent features emerge here. The first thing that stands out is the prominence of the main diagonal with most entries approximately $-1$. This comes from the construction of $\boldsymbol{\Pi}$ --- $\boldsymbol{\Pi} = \boldsymbol{\Phi}_1 - \boldsymbol{I_m}$ from equation \ref{eq:vecm} and the fact that $\boldsymbol{\hat{\Phi_i}}$ are relatively sparse in this large system with regularization.  

Next, comparing $\hat{\Pi}$ across periods shows changing structure in the long-run equilibria. The most salient change is observed in the center rows of the $\hat{\boldsymbol{\Pi}}$ matrices which correspond to the hog equations. In the \textit{Pre} and \textit{Post1} periods, there are several bright squares throughout the center columns, highlighting that several hog markets appear in many of the hog reduced-form, long-run relationships. Some of the most prominent provinces indicated by bright purple columns in the center-most submatrix are as follows: \textit{Pre}) Hubei, Neimenggu, and Shaanxi; \textit{Post1}) Guangdong, Heilongjiang, Jiangxi, Liaoning, Neimenggu. Then, in \textit{Post2}, the hog rows of $\hat{\boldsymbol{\Pi}}$ become much more concentrated with non-zero elements almost exclusively on Heilongjiang and Xinjiang  columns. 

Focusing next on the piglet and pork price rows, we see less dramatic change across the periods --- all three periods feature similar sparsity on these rows. For the piglet equations, most off-diagonal non-zero elements come from the piglet columns, with a few influences from hog and pork columns lighting up. In the pork price rows, we see a move toward concentration in \textit{Post1} and \textit{Post2}. In \textit{Pre}, there are non-zero off-diagonal elements from piglet, hog, and pork columns. In \textit{Post1} and \textit{Post2}, however, the non-zero elements begin to concentrate on the hog columns.

\subsection{Short-Run Effects}
Next we discuss the estimated $\hat{\Gamma}$ matrices in the right panel of Figure \ref{fig:var-coef}. A few features stand out. First, across all three periods we see brighter coefficients in the upper left sub-matrix that represents the piglet-to-piglet short-run effects. That suggests that the short-term spatial price relationships among the piglet prices are the strongest compared to hog-to-hog and pork-to-pig price relationships. This is likely due to the fact that piglet prices are the most volatile during all periods, as evidenced by the width of the bands and volatility over time in \ref{fig:meansd}. Comparing the piglet-to-piglet relationships across periods, we see a shift to positive and momentum short-run effects (more purple) in \textit{Post2} compared to more mean-reverting short-run effects (more red) in \textit{Pre} and \textit{Post1}. 

We also see in the \textit{Pre} period there are generally more short-run coefficients lighting up than in either \textit{Post1} or \textit{Post2} periods, even though \textit{Post1} is the most volatile period over which we fit our models (i.e., a period of great reorganization as provinces recover from the ban period). Producers had to rebuild sow herds and governments invested heavily in disease mitigation and prevention so further outbreaks would not cause similar damage.\footnote{One major action taken by provincial governments to improve disease management is to invest in large-scale, modernized hog farms post the ASF. In appendix \ref{apx:maps}, we draw a map to show the province-level expansion in large-farm production in 2020.} 

During \textit{Post1}, we also see a bright stripe of positive (purple) coefficients of piglet price in Yunnan province on all hog price equations. Yunnan is an important producer and exporter of hogs (see Figure \ref{fig:hogout}). This could be because hog prices were greatly influenced by producers' need to build back herds after the ASF outbreak. This short-run effect goes away in \textit{Post2}, indicating it was a temporary dynamic as the shock worked itself out to a new equilibrium. Apart from that, there does not seem to be major short-run impacts of any of prices on hog prices, since we see the center block of rows representing the hog equations be mostly absent from bright squares. 

Third, during the \textit{Pre} period we see many short-run effects from piglet, hog, and pork influencing pork prices, as evidenced by many colored squares in the lower third of the $\hat{\Gamma}$ matrix. However, in \textit{Post1} and \textit{Post2}, the whole area is \textit{whiter}, indicating there are fewer short-run effects impacting pork prices. This change could be attributed to multiple policy interventions aimed at stabilizing pork prices after the ASF, making pork prices less responsive to piglet and hog prices.

\subsection{Joint Impulse Response Function Analysis}
We impose three sets of one standard-deviation shocks to subsets of provincial prices in the \textit{Pre} period (i.e., the normal-time period) to show how the shocks are transmitted across space and up and down supply chain levels. Bootstrapped means of the standard deviations used for shocks in this section are found in \autoref{apx:stddev}. 

Showing all 81 JIRFs for each shock is impractical and makes it difficult to infer insights on spatial-plus-vertical price transmission.\footnote{Appendix \ref{apx:SichuanJIRF} displays eight-horizon JIRFs of one example province in the typical IRF style.} We instead use the following method to visualize a large number of JIRFs: For each shock, we display color-coded maps showing the mean response and its significance by each province for the first, third, and fifth horizon, respectively. Color shows the magnitude of response, and gray indicates the response was not statistically significant with 500 bootstrapped samples (see Appendix \ref{apx:bootstrap} for details on bootstrapping). White indicates provinces excluded from the analysis.

In figures \ref{fig:all-hog}, \ref{fig:all-pork}, and \ref{fig:all-piglet} for easier interpretation we use a $3 \times 3$ panel of maps to present bootstrap results of each shock. The first row contains the first, third, and fifth horizon responses of each provincial piglet price. The second and third rows show the responses of provincial prices for hogs and pork, respectively. Within each of the three figures, we keep the legend the same so that we can easily compare across commodities and horizons, but the legends differ across the three figures because the sizes of shocks and responses lie in different ranges. 

\subsubsection{Shocking all Hog Prices}
\begin{landscape}
\begin{figure}[!ht]
    \centering 
    \includegraphics[width=0.9\textwidth]{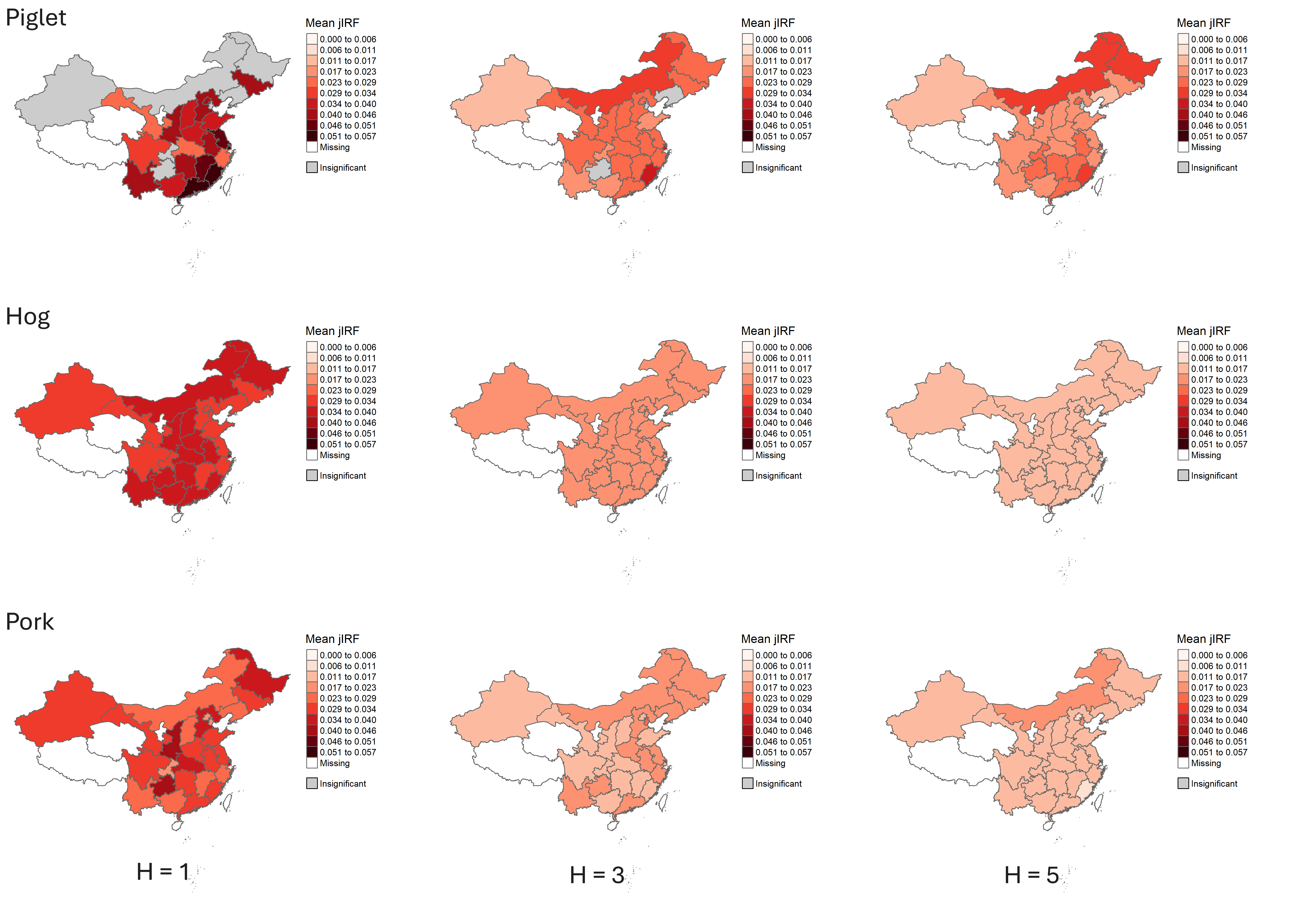}
    \caption{JIRFs of a One Standard-Deviation Shock to All Hog Prices 
    \smallskip \\ \footnotesize \textit{Note}: JIRFs for a one standard-deviation shock applied to the estimated Pre-period VECM.} 
    \label{fig:all-hog}
\end{figure}
\end{landscape}

Figure \ref{fig:all-hog} shows the JIRF results for piglet, hog, and pork prices in each province from a one standard-deviation price shock (roughly 3.5\% increase in hog prices). A few patterns catch attention. First, consider the H = 1 horizon responses of hog prices found in the second row and first column. We see fairly uniform responses across provinces to the one standard deviation increase in hog prices with prices increasing between 0.029 to 0.040 (two color bins on the map) and being statistically significant. The highly similar responses across space echo the fact that the hog markets are more spatially integrated than the piglet and pork markets in the \textit{Pre} period.\footnote{To see that hog market is more spatially integrated, we conduct pairwise cointegration tests for the \textit{Pre} period. Out of $27 \times \frac{26}{2}=351$ potential cointegrating relationships for 27 provinces, hog has 273, piglet has 187, and pork has 251. Also see the tight bands of the hog series in Figure \ref{fig:meansd}.}   

Next, consider H = 1 responses of pork prices found in the third row and first column. The pork market is less spatially integrated than the hog market (see the less tight bands of the hog series in Figure \ref{fig:meansd}), and more spatial heterogeneity is shown in the responses. Still, all responses are statistically significant and fall in the range of 0.017-0.046. Most provinces show a smaller pork price response than hog prices, but Shaanxi and Guizhou have even stronger pork price responses than the strongest hog price responses (see \autoref{fig:hogout} for names of provinces). Importantly, the heterogeneity is not clustered with any obvious regional pattern (e.g., not that all southern provinces have weaker responses). 

The H = 1 responses of piglet prices in the first row and first column show interesting contrast. The piglet market is the least spatially integrated with little inter-province trade of piglets (see the wide bands of the piglet series in Figure \ref{fig:meansd}). This structural feature is reflected in the large spatial heterogeneity of H = 1 responses of piglet prices to the shock. We see responses ranging from zero (insignificant) to 0.057. The heterogeneity does not seem to follow any particular regional pattern; e.g., Jilin's response is among the strongest, while it's neighbors to the north and south are insignificant. We do see a cluster of strongest responses in the corridor defined by Guangdong, Fujian, and Jiangxi, a cluster of insignificant responses in Chongqing, and Guizhou, and insignificant responses in most of the Northern provinces. Put together, the discussion of the H = 1 responses across commodities highlights the different degrees of spatial heterogeneity in responses to a shock in the hog market. 

Now we discuss the rate of decay and spatial heterogeneity visible in the responses at horizons 3 and 5 in figure \ref{fig:all-hog}. In row 2, we see that in addition to a fairly uniform response over space, the response to the hog price shock decays at a similar rate for all provinces. The response at H = 3 is 0.006 to 0.012 smaller than the response at H = 1; though, we see that provinces with a stronger H = 1 response decayed a bit faster since all provinces have similar responses at H = 3. Then, the rate of decay is similar, but slightly smaller, for the H = 5 responses. 

Row 3 shows dynamics in pork price responses to the shock. There is more spatial heterogeneity in the rate of decay between H = 1 and H = 3 responses; we see a lot of spatial variation in responses at H  = 1, but by H = 3, the responses become similar. As we saw with hog prices, this indicates the provinces that experienced a larger initial shock have a H = 3 shock decayed faster than provinces with a weaker H = 1 response. There is not much decay in pork prices from H = 3 to H = 5, with all but a few province responses remaining in the same color bin.

Row 1 shows the dynamics of piglet price responses. Unlike hog and even more than pork price responses, at H = 3 and 5, we still see quite a bit of spatial heterogeneity in piglet price responses. Some provinces like Fujian, Jiangxi, and Jiangsu have responses that decay across H = 1 to 3, while some other provinces show growing piglet price responses over horizons H = 1 to 3. For example, Chongqing and Heilongjiang  have insignificant price responses at H = 1, but their piglet price responses become significant at H = 3 and become even larger and significant at H = 5. 

The growing price responses are a reflection of a special feature of piglet production. Piglet demand depends on future hog prices, not immediate hog demand. As hog price increases due to the shock, for example, farmers may wait to confirm the higher expected hog prices in a few months before the decision of producing more piglets is made. This creates a lag in price responses. Besides, unlike hog and pork that have inventory ready to use any time, sows only produce about 2 litters per year, also creating a lag in price responses.\footnote{See Appendix \ref{apx:JIRFtop3} to compare shocking all hog prices with shocking hog prices in only the top three hog producing provinces. JIRFs are similar to the all hog shock, but show more spatial heterogeneity, especially in hog price responses.}

\subsubsection{Shocking all Pork Prices}
\begin{figure}[!ht]
    \centering 
    \includegraphics[width=0.9\textwidth]{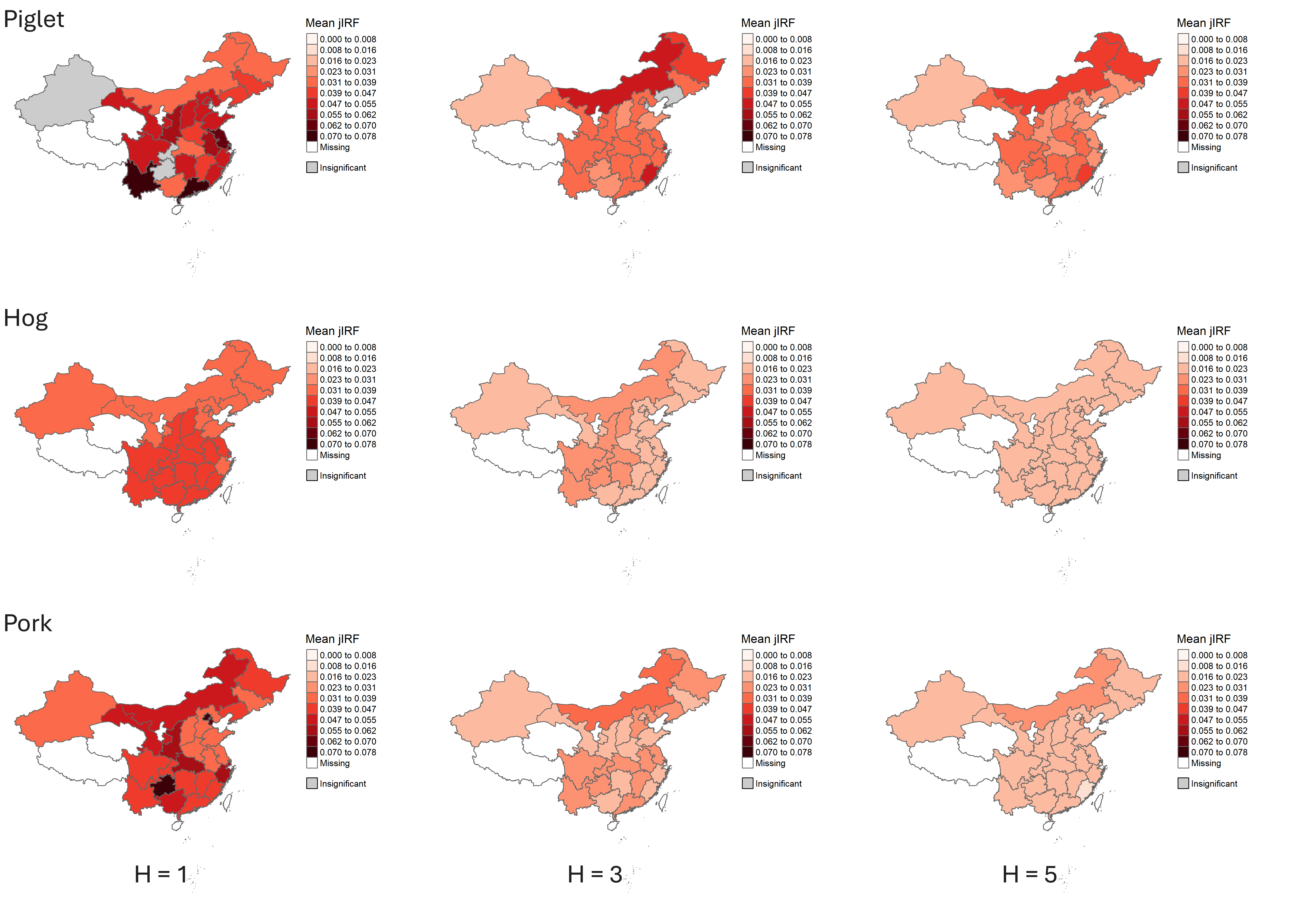}
    \caption{JIRFs of a One Standard-Deviation Shock to All Pork Prices 
    \smallskip \\ \footnotesize \textit{Note}: JIRFs for a one standard-deviation shock applied to the estimated Pre-period VECM.} 
    \label{fig:all-pork}
\end{figure}

Figure \ref{fig:all-pork} shows the JIRFs for piglet, hog, and pork prices in each province from a one standard-deviation shock to all pork prices. Pork price shocks range from 3.5\% to 8.0\% across provinces.

In row 2 we see hog price responses to pork price shocks that are similar in size as the hog price responses to the hog price shock and similar in that they are fairly spatially homogeneous. At H = 1 we see our first clear spatial clustering with Southern provinces responding slightly more strongly to pork price shocks than they did for hog price shocks. Then, by H = 3 we see stronger responses clustered in Central corridor provinces, but by H = 5, all provinces have a similar response. 

Row 3 shows that pork prices have more spatial heterogeneity at H = 1 than hog price responses; at H = 1 we see Beijing, Tianjin (major urban consumers of pork) and Guizhou with the strongest responses (0.070 to 0.078), while several provinces' responses are in the range 0.031 to 0.039. We see a similar heterogeneous decay from H = 1 to 3 as we saw with hog price shocks, and by H = 5 most pork price responses are in the same color bin. 

In row 1 we see that piglet prices again have the most spatial heterogeneity in responses. Three province price responses are insignificant and two have among the highest responses of any in the system to this shock. Most provinces see responses decay over horizons 1 to 5, but Heilongjiang responses get stronger.

\subsubsection{Shocking All Piglet Prices}
\begin{figure}[!ht]
    \centering 
    \includegraphics[width=0.9\textwidth]{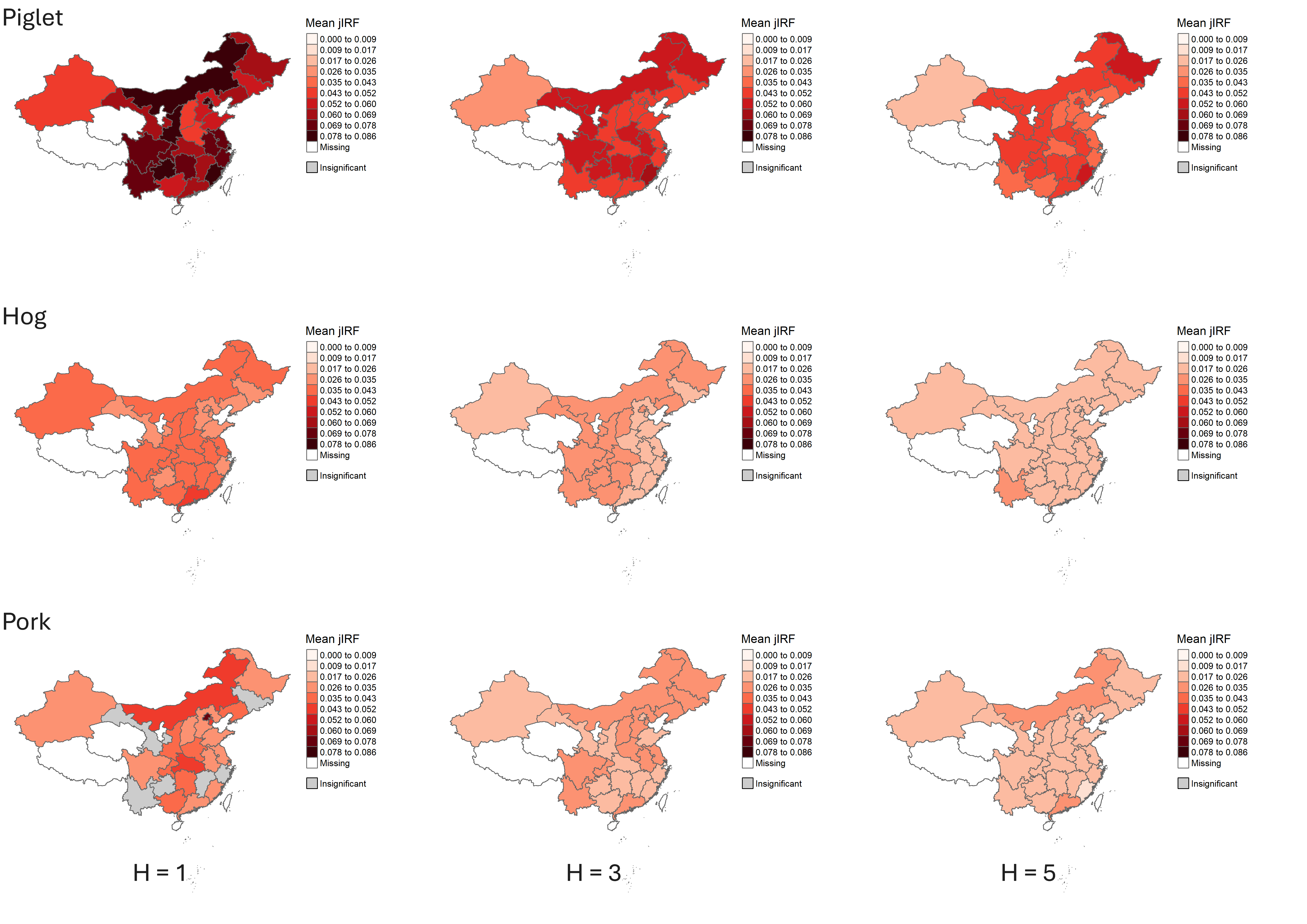}
    \caption{JIRFs of a One Standard-Deviation Shock to All Piglet Prices 
    \smallskip \\ \footnotesize \textit{Note}: JIRFs for a one standard-deviation shock applied to the estimated Pre-period VECM.} 
    \label{fig:all-piglet}
\end{figure}

Figure \ref{fig:all-piglet} shows the JIRFs for piglet, hog, and pork prices in each province from a one standard-deviation shock to all piglet prices. Piglet price shocks range from about 6\% to 10\% across provinces.   

In row 2, we see the hog price responses are fairly uniform with similar responses across the provinces in the range of 0.026 to 0.052 at H = 1, and fairly uniform rates of decay across horizons 3 and 5. 

Row 3 shows that pork price responses to piglet price shocks have considerable spatial heterogeneity. As in other shock scenarios, there is no discernible regional pattern. However, as we look at the responses across horizons, we see a different decay pattern where Yunnan, Gansu, Jiangxi, Guizhou, Jilin, and Zhejiang go from insignificant at H = 1 to significant at H = 3, and remain significant with the same level of response at H = 5 as seen at H = 3.

Row 1 shows the piglet price responses to the piglet shock. As in the other scenarios, piglet prices show the most spatial heterogeneity in responses (ranging from 0.043 to 0.086). Looking from H = 1 to H = 5, we still see substantial spatial variation in responses. Overall, responses decay. At H = 5 responses span from 0.017 to 0.052. But a few provinces like Henan see increased responses over time, again reflecting the lags in piglet production.

\section{Concluding Remarks}
We show how price transmission of shocks to multiple price series can be estimated by the JIRF applied to a large system of prices whose relationships are estimated via regularized regression in a VECM. The regularized regression approach allows rich relationships across space and along the supply chain to be uncovered that would be missed if prices are aggregated or left out to enable traditional econometric estimation. 

Applying this approach to the piglet-hog-pork markets in 27 Chinese provinces, we show that there are rich spatial and vertical relationships among these markets and that they experienced structural breaks upon the imposition and removal of the inter-province shipping ban of live hogs due to the ASF outbreak in 2018. To aid the interpretation of the price relationships uncovered by the VECM, we compute JIRFs with bootstrapped confidence intervals. The JIRF allows the researcher to shock subsets of the variables, and we explored three main scenarios of a one standard-deviation shock in the price: shocking all hog prices, shocking all pork prices, and shocking all piglet prices. We find that across all the hypothetical scenarios, hog prices have a fairly uniform response across space and also decay at similar rates. Pork and piglet prices show more spatial heterogeneity in initial responses and in the rates at which responses to shocks dissipate. Piglet prices show the most persistent responses to shocks, with some provinces showing growing responses over the horizons considered, rather than the more typical pattern where prices see a large initial response that decays over time. 

Our work has implications not only for researchers, but also for practitioners, policymakers, and stakeholders. Combining the ability to estimate high-dimensional price relationships and interpret their responses to shocks allows for greater preparation for shocks and better ways to mitigate their effects. With increasingly frequent disruptions to supply chains in agrifood and other industries, the ability to perform this kind of stress-test will become increasingly important. 

Future research could extend our work in several ways. First, newly developed debiasing methods of for machine learning models could be applied in settings where small samples are a concern. Second, our approach could be applied to a number of other contexts that are relevant to food and agriculture and beyond; e.g., in international trade, multi-supply-chain settings, or even in how pollution dissipates through a landscape. Further, other components known to impact a market could be directly modeled, whether it be energy demand, weather, or prices of substitutes or complements, for example. Third, other ML models could be used to produce estimates of price relationships that can be explored with JIRFs. Other ML models may be better suited depending on the nature of the question and the data being considered.

\newpage
\singlespacing
\setlength\bibsep{10pt}
\bibliography{biblio}


\newpage
\appendix
\setcounter{table}{0}
\setcounter{figure}{0}
\renewcommand{\thetable}{A\arabic{table}}
\renewcommand{\thefigure}{A\arabic{figure}}

\section{Hog and Pork Markets in China}\label{apx:maps}
\begin{figure}[!ht]
    \centering 
    \includegraphics[width=0.85\textwidth]{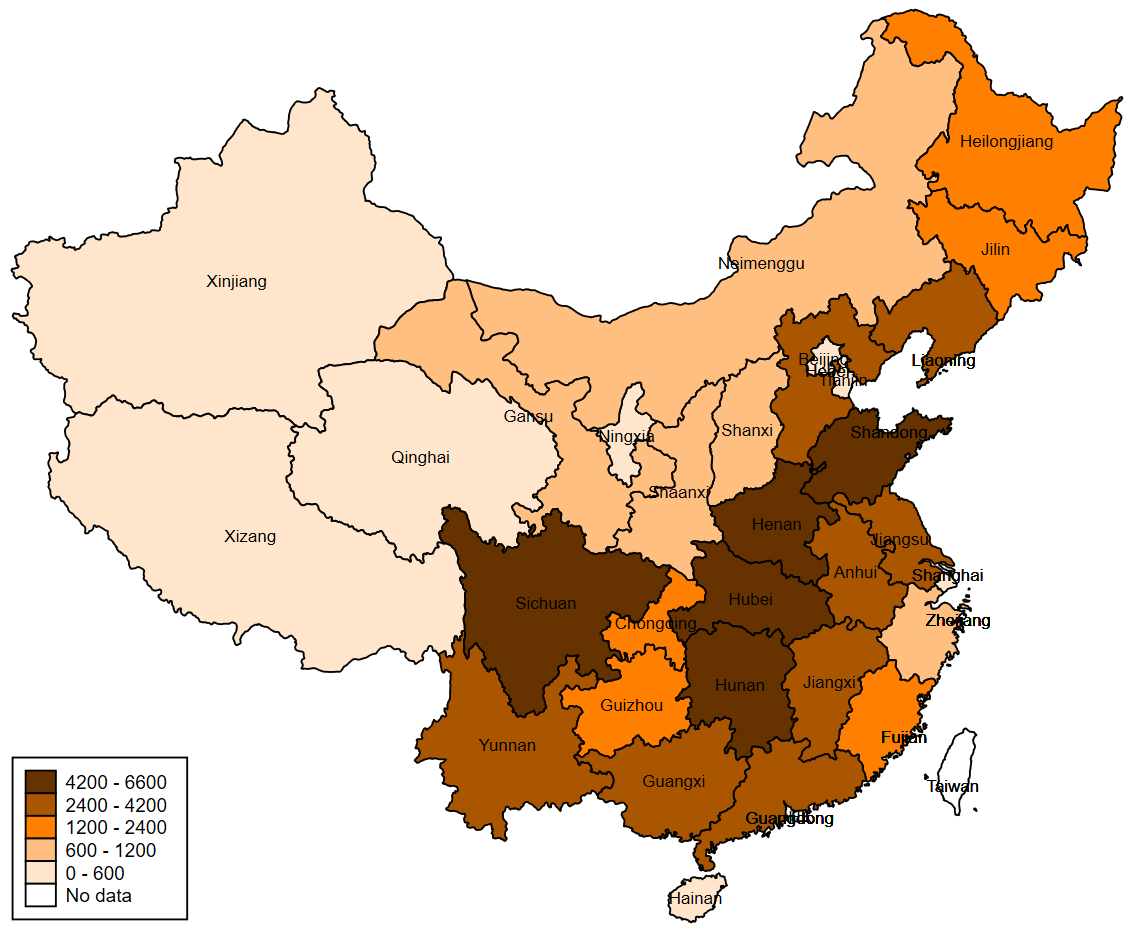}
    \caption{Hog Output by Province, China, 2017
    \smallskip \\ \footnotesize \textit{Note}: Data come from China's provincial yearbooks. The output is measured in 10,000 metric tons.} 
    \label{fig:hogout}
\end{figure}

We visualize the scale of hog production in the Chinese provinces. Figure \ref{fig:hogout} shows the distribution of hog production prior to the ASF outbreak. The leading producers are Henan, Hubei, Hunan, Sichuan, and Shandong. As mentioned, the provinces with large numbers of missing observations are all small producers of hogs and hence not harmful to exclude from our empirical analysis. 

\begin{figure}[!ht]
    \centering
\begin{subfigure}{0.65\textwidth}
     \resizebox{\textwidth}{!}{\includegraphics{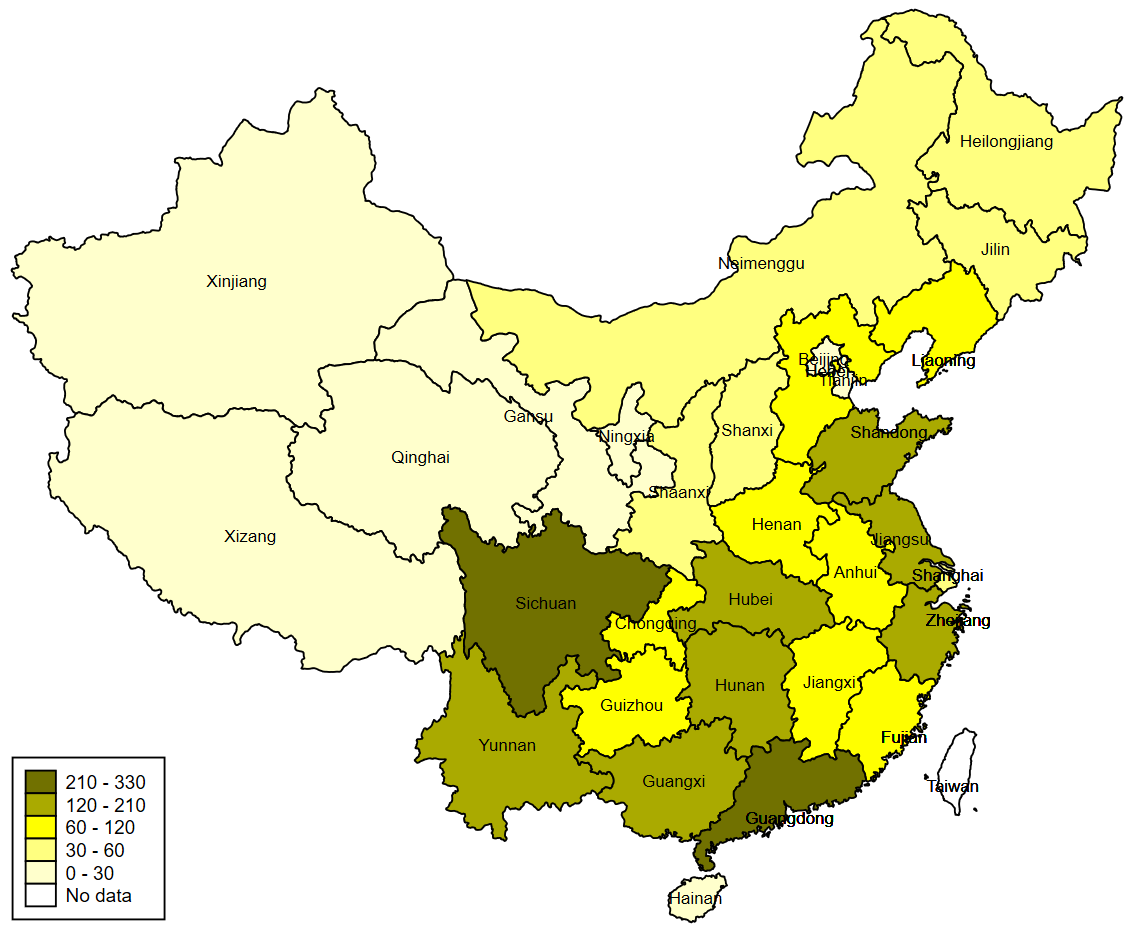}}
      \subcaption{\footnotesize Total Pork Consumption by Province}
    \end{subfigure}
\begin{subfigure}{0.65\textwidth}
     \resizebox{\textwidth}{!}{\includegraphics{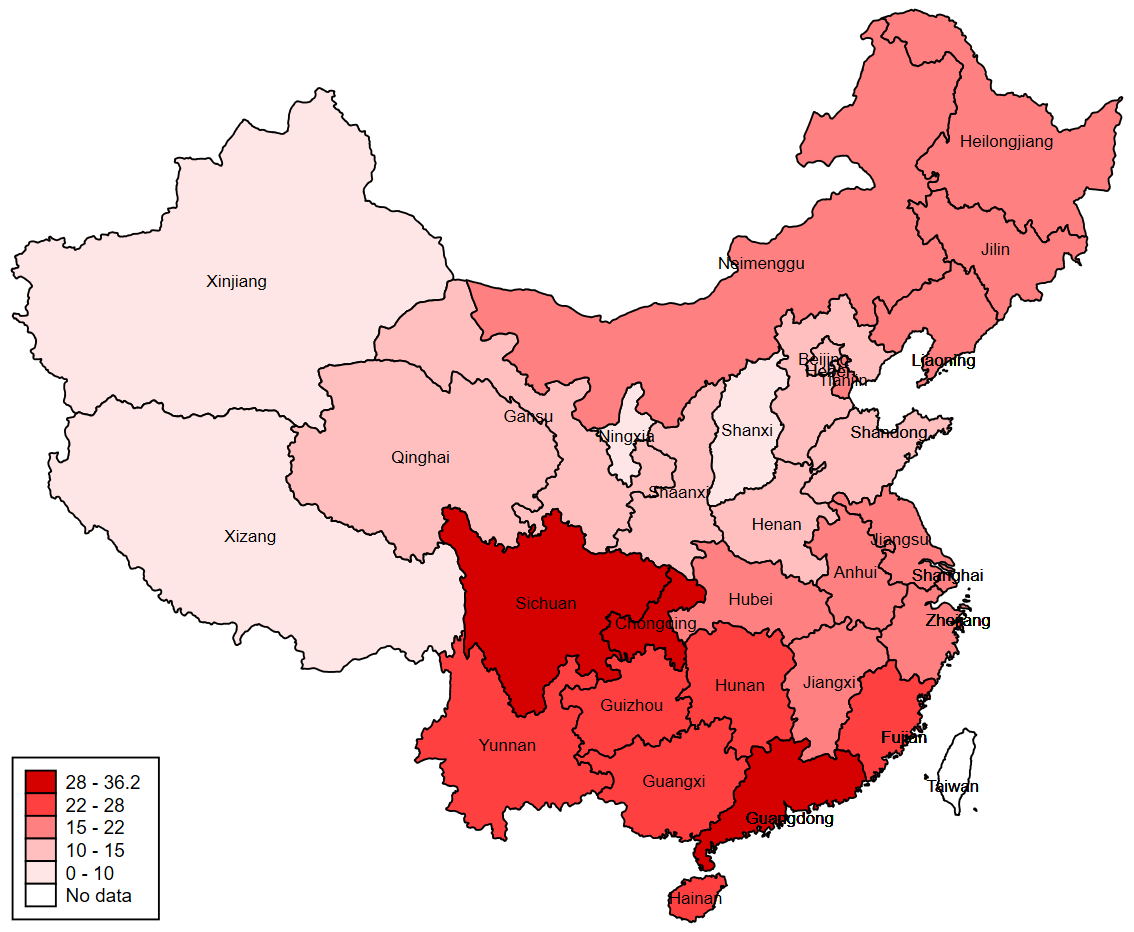}}
      \subcaption{\footnotesize Per Capita Pork Consumption by Province}
    \end{subfigure}  
\caption{Pork Consumption by Province, China, 2017
    \smallskip \\ \footnotesize \textit{Note}: Data come from China's provincial yearbooks. The total consumption is measured in 10,000 metric tons, and the per capita consumption is measured in kilograms.} 
    \label{fig:porkconsp}
\end{figure}

Figure \ref{fig:porkconsp} shows the province-level total and per capita pork consumption prior to the ASF outbreak. The four leading consumers are Guangdong, Hunan, Jiangsu, and Sichuan. We can see that the top consumer provinces do not tend to overlap with the top producer provinces. It is also worth noting the provinces with large numbers of missing observations are all small consumers of pork. 

\begin{figure}[!ht]
    \centering 
    \includegraphics[width=0.65\textwidth]{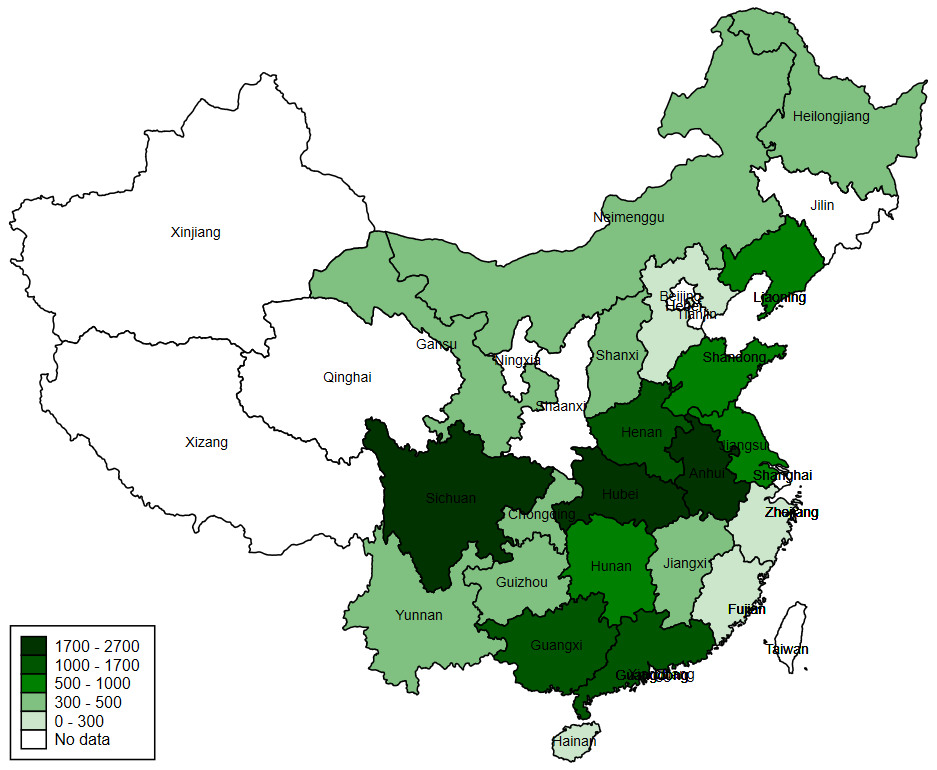}
    \caption{Hog output Expansion by Province, China, 2020
    \smallskip \\ \footnotesize \textit{Note}: Data come from \url{http://www.sdxmjingji.com/zhengwugongkai/renshixinxi/264.html} (in Chinese). The output expansion is measured in 10,000 head per year.} 
    \label{fig:hoginv}
\end{figure}

\clearpage
Figure \ref{fig:hoginv} shows the province-level incremental hog production in 2020. The four leading provinces in expanding hog production are Sichuan (26.1 million head per year), Hubei (22.0 million head per year), Anhui (17.4 million head per year), Guangdong (15.3 million head per year), and Henan (15.0 million head per year). These provinces are also leading producers before the ASF (see Figure \ref{fig:hogout}), implying more concentrated production in China.

\clearpage
\setcounter{table}{0}
\setcounter{figure}{0}
\renewcommand{\thetable}{B\arabic{table}}
\renewcommand{\thefigure}{B\arabic{figure}}

\section{JIRF Scenarios and Bootstrap Procedure}\label{apx:stddev}
We explore three main shock scenarios to generate our JIRF analyses. We shock all piglet, all hog, and all pork province prices by one standard deviation. The standard deviation is calculated by province over the \textit{Pre} period. Thus, if the $i$th province price is one of the shocked variables, the magnitude of the shock is one standard deviation 
$$\sigma_{i} = \sqrt{\frac{\sum_{t=1}^{T}(y_{i,t} - \mu_i)^2}{T}},$$ 
where $\mu_i$ is the mean of $y_{i,t}$ and the magnitude of the shock is zero if the price is not one of the shocked variables. Table \ref{tab:stdev} reports the magnitudes of one standard-deviation shocks for the JIRF simulations.  The Piglet column shows the province-level shocks in the Piglet shock scenario, and the Hog and Pork shocks are implemented similarly.

\begin{table}[!ht]
\centering\caption{One Standard-Deviation Shocks Used in JIRF Scenarios}
\begin{threeparttable} 
\begin{tabular}{rrrr}
  \toprule
 & Piglet & Hog & Pork \\ 
  \midrule
  Anhui & 0.0765 & 0.0368 & 0.0406 \\ 
  Beijing & 0.0938 & 0.0350 & 0.0827 \\ 
  Chongqing & 0.0843 & 0.0350 & 0.0490 \\ 
  Fujian & 0.0896 & 0.0369 & 0.0445 \\ 
  Gansu & 0.0681 & 0.0317 & 0.0518 \\ 
  Guangdong & 0.0730 & 0.0373 & 0.0410 \\ 
  Guangxi & 0.0644 & 0.0352 & 0.0547 \\ 
  Guizhou & 0.0929 & 0.0355 & 0.0832 \\ 
  Hebei & 0.0597 & 0.0356 & 0.0361 \\ 
  Heilongjiang & 0.0688 & 0.0384 & 0.0447 \\ 
  Henan & 0.0542 & \textbf{0.0364} & 0.0406 \\ 
  Hubei & 0.0714 & 0.0357 & 0.0635 \\ 
  Hunan & 0.0762 & \textbf{0.0373} & 0.0469 \\ 
  Jiangsu & 0.0740 & 0.0328 & 0.0378 \\ 
  Jiangxi & 0.0717 & 0.0348 & 0.0500 \\ 
  Jilin & 0.0620 & 0.0333 & 0.0379 \\ 
  Liaoning & 0.0678 & 0.0336 & 0.0440 \\ 
  Neimenggu & 0.0955 & 0.0395 & 0.0604 \\ 
  Shandong & 0.0556 & 0.0342 & 0.0340 \\ 
  Shanghai & 0.0988 & 0.0456 & 0.0642 \\ 
  Shaanxi & 0.0846 & 0.0354 & 0.0589 \\ 
  Shanxi & 0.0531 & 0.0370 & 0.0343 \\ 
  Sichuan & 0.0809 & \textbf{0.0344} & 0.0445 \\ 
  Tianjin & 0.0950 & 0.0355 & 0.0708 \\ 
  Xinjiang & 0.0546 & 0.0330 & 0.0365 \\ 
  Yunnan & 0.0783 & 0.0374 & 0.0482 \\ 
  Zhejiang & 0.0818 & 0.0313 & 0.0597 \\ 
   \bottomrule
\end{tabular}

\begin{tablenotes}
\footnotesize{
\textit{Note}: Shocks used for scenarios in the JIRF analysis are one standard deviations to the following subsets of prices: all piglet prices, all hog prices, all pork prices, and top three hog producer prices. Table shows the value of these one standard-deviation shocks imposed. Three bold numbers indicate the top three hog producing provinces.}
\end{tablenotes}
\end{threeparttable}
\label{tab:stdev}
\end{table}

\subsection{Bootstrap Procedures}\label{apx:bootstrap}
To assess the sampling variability of our VECM estimates, we implement a residual-based bootstrap procedure, closely following standard time series resampling techniques \citep{berkowitz2000recent}. This procedure generates synthetic datasets by resampling the residuals from the estimated VAR model and reconstructing the time series using the estimated dynamics.

 Consider the estimated VAR(2) model on which our VECM is based,
\[
\mathbf{Y}_{t} = \mathbf{c} + \mathbf{\Phi}_{1} \mathbf{Y}_{t-1} + \mathbf{\Phi}_{2} \mathbf{Y}_{t-2} + \boldsymbol{\epsilon}_{t},
\]
where $\mathbf{Y}_t \in \mathbb{R}^m$ is the vector of endogenous variables, $\mathbf{c}$ is a vector of intercept terms, $\mathbf{\Phi}_{1}$ and $\mathbf{\Phi}_{2}$ are $m \times m$ coefficient matrices, and $\boldsymbol{\epsilon}_t$ are iid innovations with mean zero and covariance matrix $\Sigma$. Let $\hat{\boldsymbol{\epsilon}}_t$ denote the residuals from the estimated model. These residuals are first centered to ensure zero mean: $\tilde{\boldsymbol{\epsilon}}_t = \hat{\boldsymbol{\epsilon}}_t - \bar{\boldsymbol{\epsilon}}$, with $\bar{\boldsymbol{\epsilon}} = \frac{1}{T - 2} \sum_{t = 3}^T \hat{\boldsymbol{\epsilon}}_t$. The bootstrap sample $\{\mathbf{Y}^*_t\}$ is initialized using the first two values of the original series, i.e., $\mathbf{Y}^*_t = \mathbf{Y}_t$ for $t = 1, 2$. For $t > 2$, a residual $\tilde{\boldsymbol{\epsilon}}^*_t$ is drawn with replacement from the centered residuals $\{\tilde{\boldsymbol{\epsilon}}_s\}_{s = 3}^T$, and the synthetic data is recursively generated as
\[
\mathbf{Y}^*_t = \hat{\mathbf{c}} + \hat{\mathbf{\Phi}}_{1} \mathbf{Y}^*_{t-1} + \hat{\mathbf{\Phi}}_{2} \mathbf{Y}^*_{t-2} + \tilde{\boldsymbol{\epsilon}}^*_t.
\]

This process preserves the estimated dynamic structure of the VAR(2) model, while introducing variation through resampled innovations. Repeating this procedure $B$ times yields an empirical distribution  $\hat{\mathbf{\Phi}}_{1}$ and $\hat{\mathbf{\Phi}}_{2}$. We then transform the estimated VAR coefficients  into a VMA process to obtain Equation \ref{eq:VMA} and compute the JIRF using Equation \ref{eq:JIRF} to get the empirical distribution of the JIRF.

\subsection{Classical Impulse Response-Style Figure}\label{apx:SichuanJIRF}
\begin{figure}[!h]
    \centering 
    \includegraphics[width=1.0\textwidth]{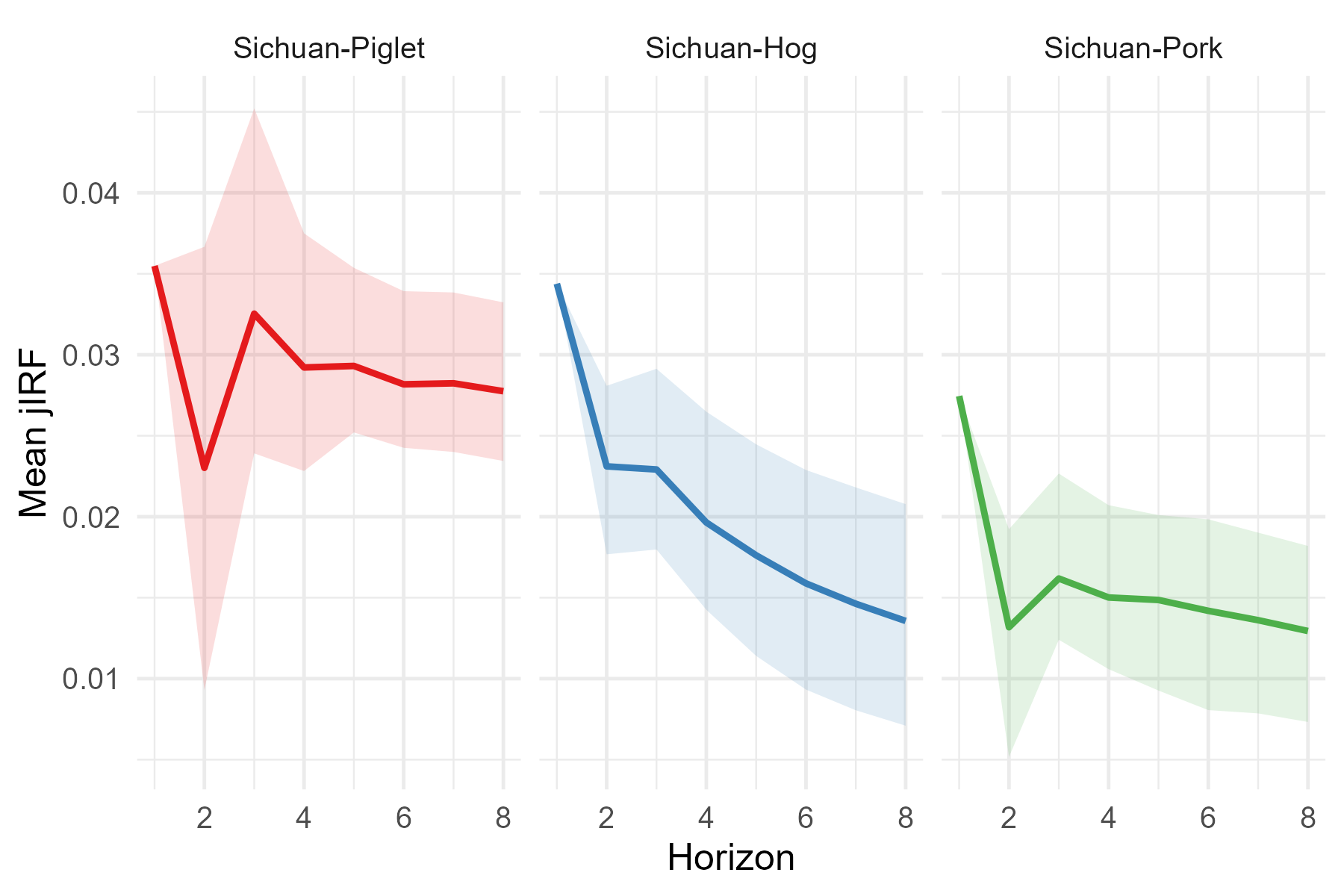}
    \caption{Sichuan JIRFs of a One Standard-Deviation Shock to All Hog Producing Province Prices 
    \smallskip \\ \footnotesize \textit{Note}: Sichuan JIRFs for a one standard-deviation shock applied to the estimated Pre-period VECM.}     \label{fig:classic_irf}
\end{figure}

For comparison to the classical impulse response-style graphs produced from SVAR, we show in figure \ref{fig:classic_irf} the JIRF results of Sichuan piglet, hog, and pork prices to shocking all hog prices. 

\subsection{JIRF of Shocking the Top Three Hog Producers}\label{apx:JIRFtop3}
\begin{figure}[!h]
    \centering 
    \includegraphics[width=1.0\textwidth]{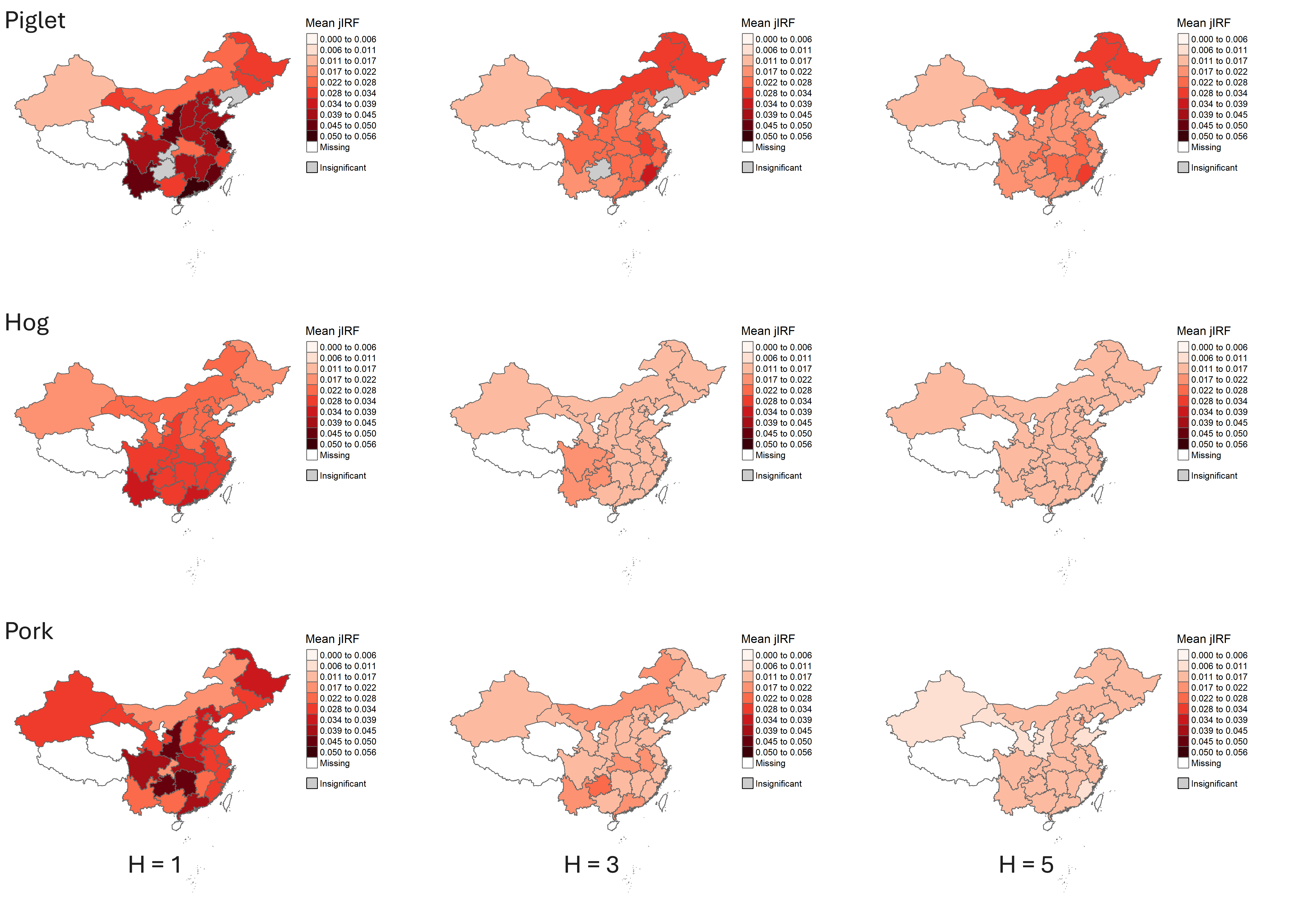}
    \caption{JIRFs of a One Standard-Deviation Shock to Top Hog Producing Province Prices 
    \smallskip \\ \footnotesize \textit{Note}: JIRFs for a one standard-deviation shock applied to the estimated Pre-period VECM. Confidence intervals computed from 500 bootstrapped samples.}     \label{fig:tophog}
\end{figure}

In figure \ref{fig:tophog}, we show the JIRFs when only shocking the hog prices of the top 3 hog producing provinces, Henan, Hunan, and Sichuan. The results can be compared with the JIRFs produced by shocking all hog province prices to reveal similar effects (see \autoref{fig:all-hog}), but more spatial heterogeneity in responses when we shock a smaller number of price series, compared to a more uniform shock applied to a larger subset of prices. 

\clearpage
\setcounter{table}{0}
\setcounter{figure}{0}
\renewcommand{\thetable}{O\arabic{table}}
\renewcommand{\thefigure}{O\arabic{figure}}
\renewcommand{\thesubsection}{O.1}

\section{Online Appendix}\label{apx:online}
This online appendix reports detailed model coefficient estimates. Readers can find commodity names and province names along the horizontal and vertical axes in the following figures.  

\begin{figure}[!h]
    \centering 
    \includegraphics[width=1.0\textwidth]{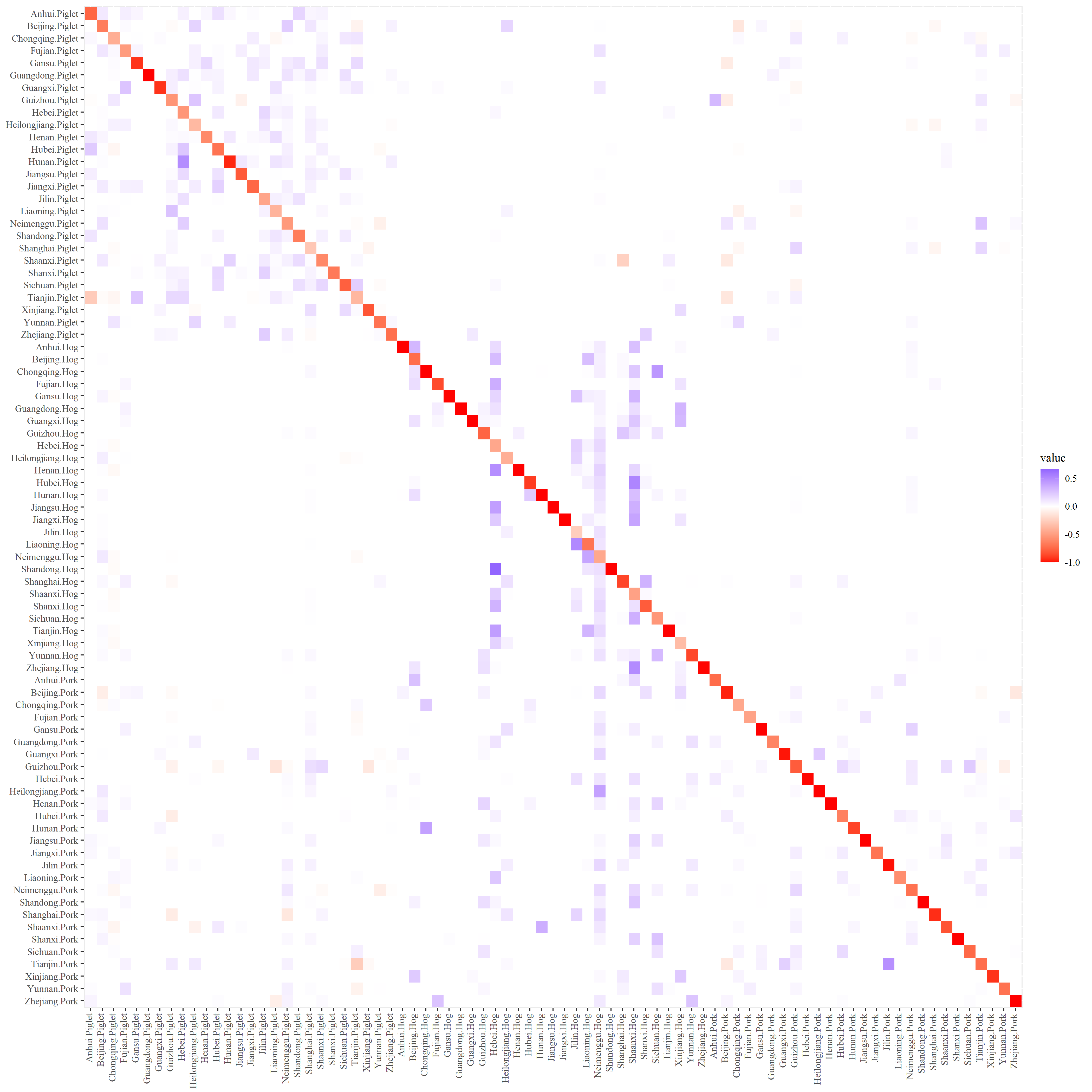}
    \caption{$\Pi$ Matrix, \textit{Pre} Period
    \smallskip \\ \footnotesize \textit{Note}: Estimated $\Pi$ matrix from equation \ref{eq:vecm}}     \label{fig:prepi}
\end{figure}

\begin{figure}[!ht]
    \centering 
    \includegraphics[width=1.0\textwidth]{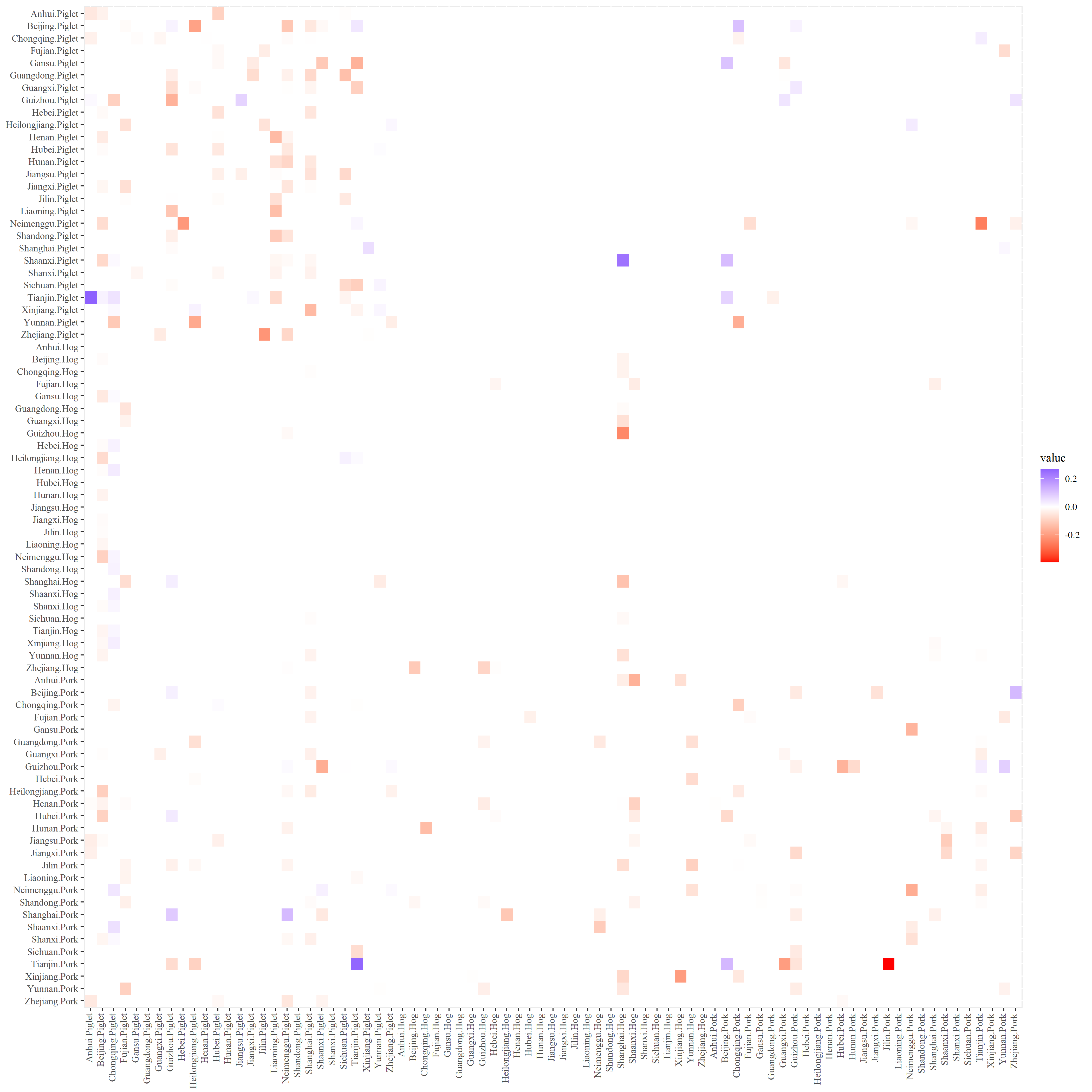}
    \caption{$\Gamma$ Matrix, \textit{Pre} Period
    \smallskip \\ \footnotesize \textit{Note}: Estimated $\Gamma$ matrix from equation \ref{eq:vecm}} 
    \label{fig:pregama}
\end{figure}

\begin{figure}[!ht]
    \centering 
    \includegraphics[width=1.0\textwidth]{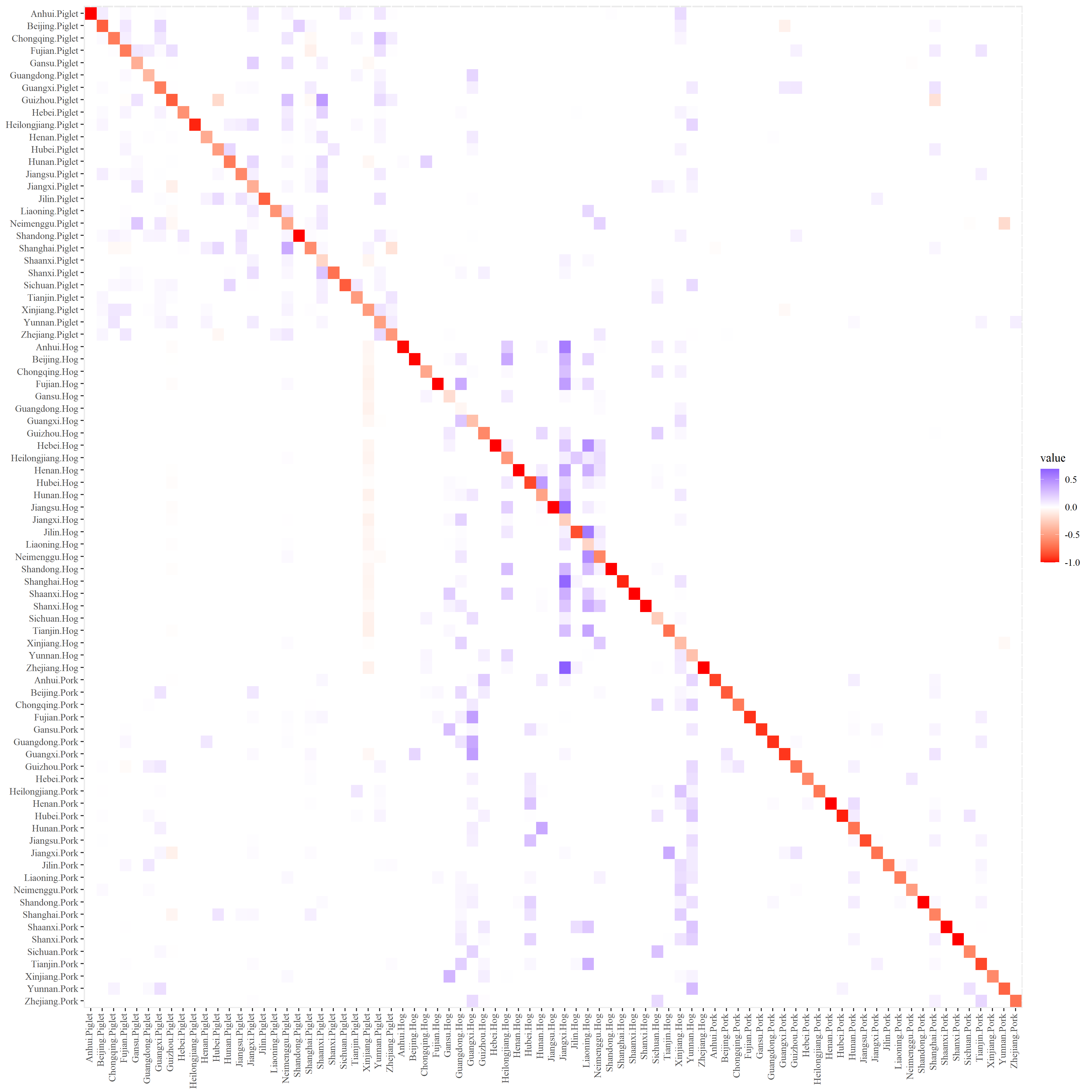}
    \caption{$\Pi$ Matrix, \textit{Post1} Period
    \smallskip \\ \footnotesize \textit{Note}: Estimated $\Pi$ matrix from equation \ref{eq:vecm}} 
    \label{fig:post1pi}
\end{figure}

\begin{figure}[!ht]
    \centering 
    \includegraphics[width=1.0\textwidth]{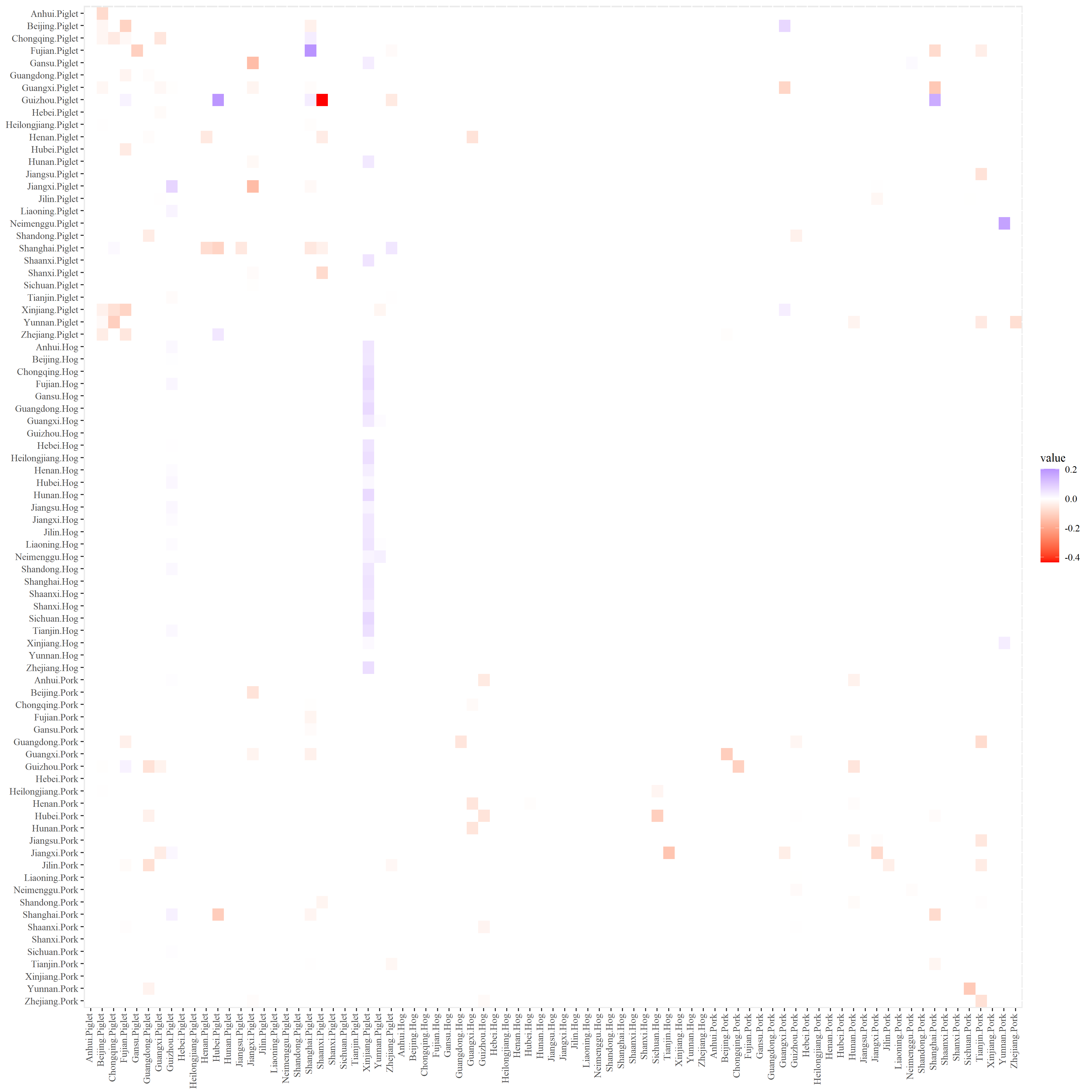}
    \caption{$\Gamma$ Matrix, \textit{Post1} Period
    \smallskip \\ \footnotesize \textit{Note}: Estimated $\Gamma$ matrix from equation \ref{eq:vecm}} 
    \label{fig:post1gamma}
\end{figure}

\begin{figure}[!ht]
    \centering 
    \includegraphics[width=1.0\textwidth]{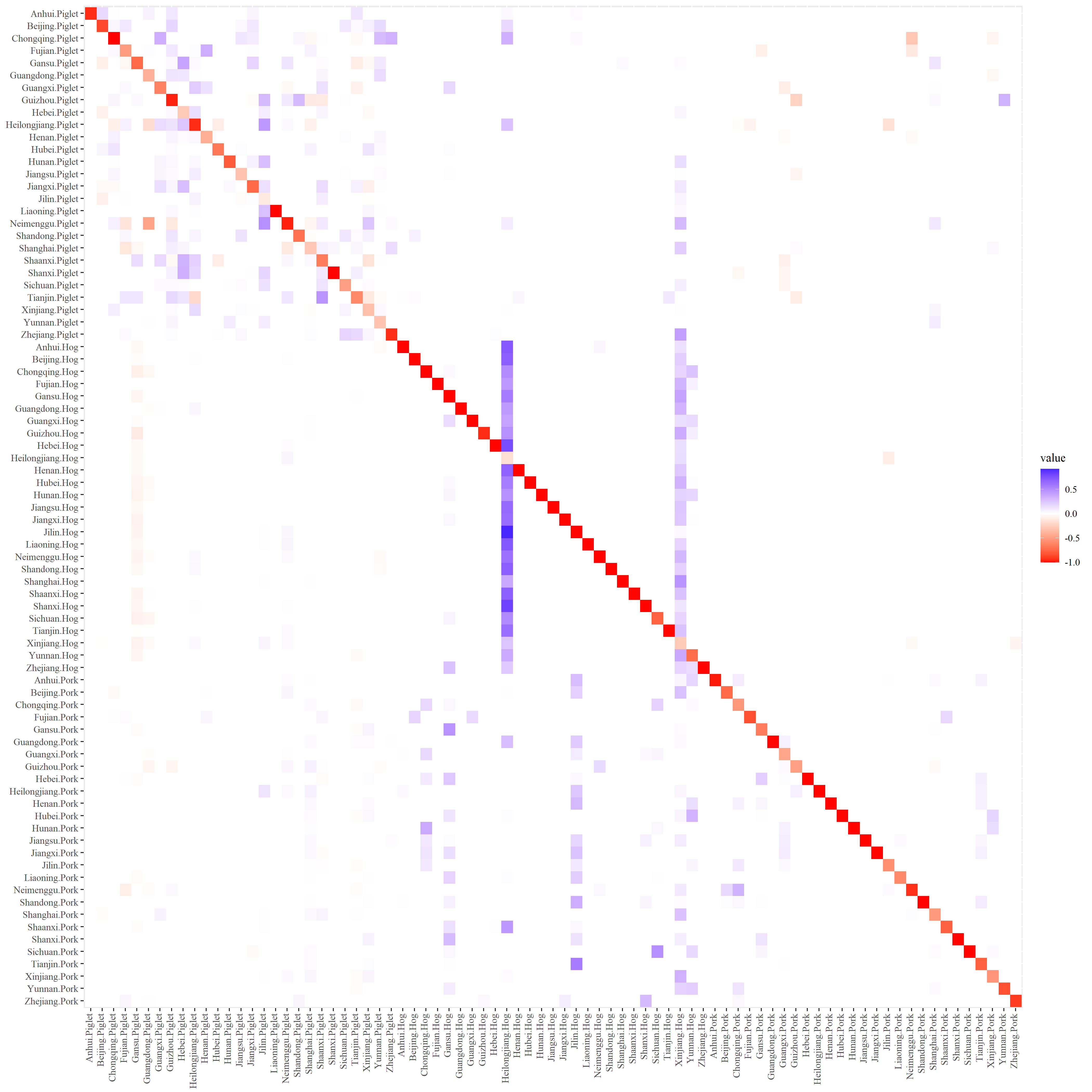}
    \caption{$\Pi$ Matrix, \textit{Post2} Period
    \smallskip \\ \footnotesize \textit{Note}: Estimated $\Pi$ matrix from equation \ref{eq:vecm}} 
    \label{fig:post2pi}
\end{figure}

\begin{figure}[!ht]
    \centering 
    \includegraphics[width=1.0\textwidth]{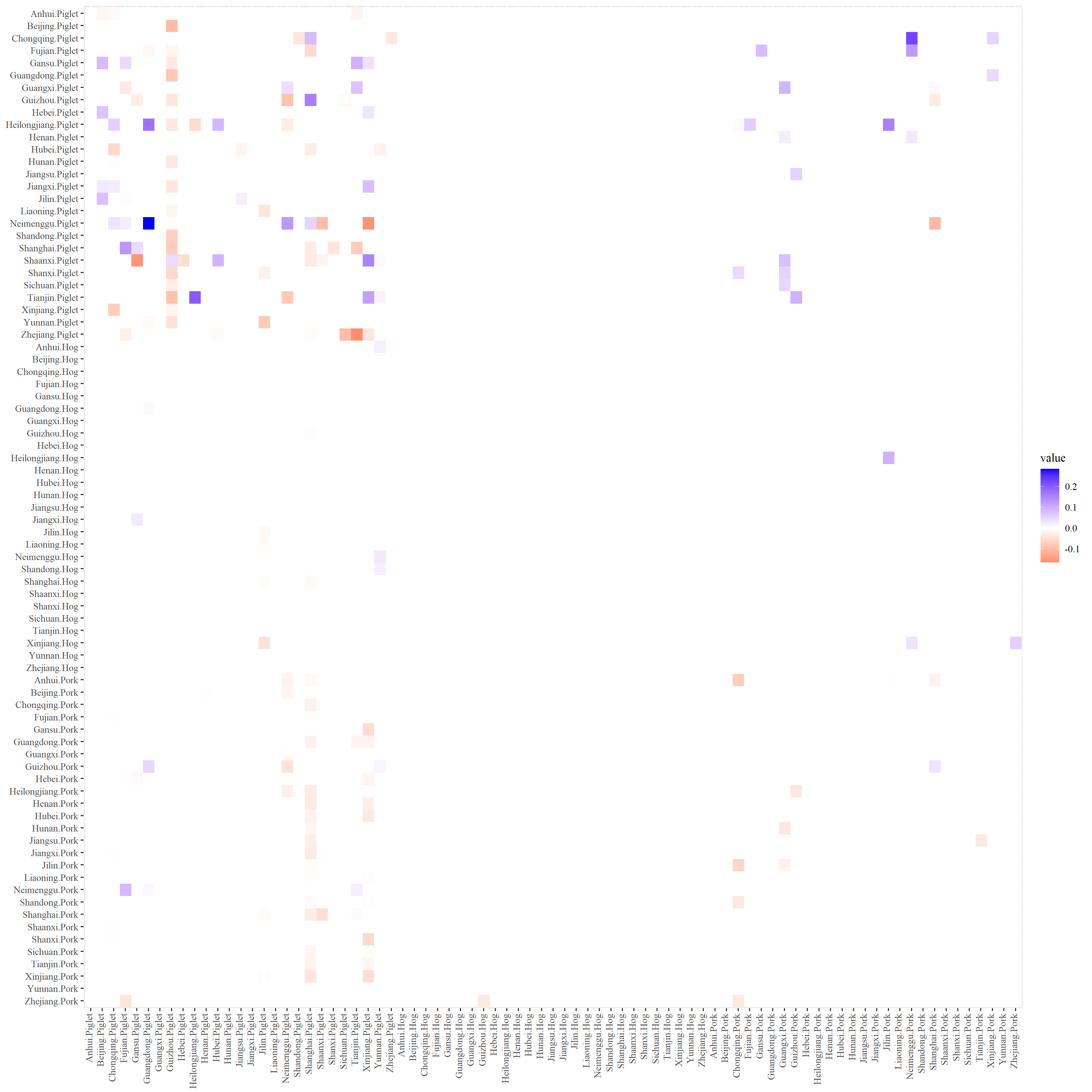}
    \caption{$\Gamma$ Matrix, \textit{Post2} Period
    \smallskip \\ \footnotesize \textit{Note}: Estimated $\Gamma$ matrix from equation \ref{eq:vecm}} 
    \label{fig:post2gamma}
\end{figure}

\clearpage
\subsection{Rank of $\Pi$}\label{apx:rank}
The rank of the estimated $\boldsymbol{\hat{\Pi}}$ matrix determines the number of linearly independent equilibrium relationships among the variables in the model. As mentioned, if we could identify an economically meaningful factorization of $\boldsymbol{\hat{\Pi}} =\boldsymbol{\hat{\alpha} \hat{\beta'}}$, we could examine the degree of spatial and vertical price linkages directly through the $\boldsymbol{\hat{\beta'}}$ matrix. This is not possible, however, as discussed. We will therefore take a different approach. 

To examine the nature of spatial price linkages specifically, we fit a penalized least squares model for each period and for each price type separately. For example, we fit one model to all piglet prices from all twenty-seven provinces in our sample. Then we compare the effective rank of $\boldsymbol{\hat{\Pi}}$ across periods to determine the nature of the complexity of spatial linkages among piglet prices across periods. We do the same for the hog prices and the pork prices. 

The results of the effective rank calculations are shown in Table \ref{tab:ranks}. For the piglet, hog, and pork price models, full rank would be 27. We find that for each period, the effective rank within a type is nearly full rank. For example, the rank of $\boldsymbol{\hat{\Pi}}$ for piglet prices is 23.8, 23.7, and 22.9 across the Pre, \textit{Post1}, and \textit{Post2} periods, respectively. For hog prices, these values are 23.6, 22.2, and 22.4, respectively, and for pork prices the effective ranks are 24.3, 23.5, and 22.1, respectively.

\begin{table}[!ht]
    \centering\caption{Estimated Effective Ranks by Commodity and Period}
\begin{tabular}{lccc}
\hline
&\multicolumn{3}{c}{Subperiod} \\
 & Pre & Post1 & Post2 \\ 
\hline
Piglet & 23.80 & 23.69 & 22.92 \\ 
Hog & 23.56 & 22.23 & 22.42 \\ 
Pork & 24.28 & 23.51 & 22.13 \\ 
\hline
Sum & 70.41 & 68.22 & 69.92 \\ 
\hline
\\
Piglet, hog, and pork & 73.44 & 71.95 & 71.78 \\ 
\hline
\end{tabular}

\begin{tablenotes}
\item \footnotesize{\textit{Note}: Authors' calculation.}
\end{tablenotes}
\label{tab:ranks}
\end{table}

The large effective ranks are supportive that the VECM is an appropriate model specification, i.e., that $r >0$. Next, the large effective ranks within the spatial-only systems (i.e., the upper panel in the table) indicates each of the three spatial price systems shows spatial linkages and the nature of the linkages are complex. We say they are complex because the spatial equilibrium are made up of a large number of linearly independent spatial equilibria.  We would contrast this against 'simple' spatial linkages where there is one or a small number of spatial equilibria present. Note that a large number of equilibrium relationships should not be confused as 'more spatially integrated'. It simply reveals the nature of the spatial network, but does not indicate whether linkages are strong or weak. 

It is also difficult to confidently interpret the effective ranks across periods because we do not have any way to statistically compare the effective rank across time-periods. We note that for piglet, hog, and pork, respectively, effective ranks that are slightly lower in \textit{Post1} and \textit{Post2} than in the \textit{Pre} period.

\end{document}